\documentclass[a4paper]{spie}

\pdfoutput=1

\usepackage{amsmath,amsfonts,amssymb}
\usepackage{graphicx}
\usepackage[colorlinks=true, allcolors=blue]{hyperref}
\usepackage{siunitx}
    \DeclareSIUnit\week{wk}
    \DeclareSIUnit\year{yr}
\usepackage{subcaption}
\usepackage{aas_macros}
\usepackage[left=1.925cm,top=2.94cm,right=1.925cm,bottom=4.94cm,bindingoffset=0cm]{geometry}
\usepackage{floatrow}
\newcommand{\hlinep}{\noalign{\smallskip} \hline \noalign{\smallskip}}

\makeatletter
\def\bstctlcite{\@ifnextchar[{\@bstctlcite}{\@bstctlcite[@auxout]}}
\def\@bstctlcite[#1]#2{\@bsphack
  \@for\@citeb:=#2\do{%
    \edef\@citeb{\expandafter\@firstofone\@citeb}%
    \if@filesw\immediate\write\csname #1\endcsname{\string\citation{\@citeb}}\fi}
  \@esphack}
\makeatother

\title{Echidna Mark II: one giant leap for `tilting spine' \\
       fibre positioning technology}

\author[a,b,c]{James Gilbert*}
\author[a,d]{Gavin Dalton}
\affil[a]{University of Oxford Department of Physics, Keble Road, Oxford OX1 3RH, UK}
\affil[b]{Australian Astronomical Observatory, 105 Delhi Road, North Ryde, NSW 2113, Australia}
\affil[c]{Research School of Astronomy \& Astrophysics, The Australian National University, Cotter Road, Weston Creek, ACT 2611, Australia}
\affil[d]{RAL Space, STFC Rutherford Appleton Laboratory, Didcot OX11 0QX, UK}

\authorinfo{*james.gilbert@physics.ox.ac.uk}

\begin{document} 
\maketitle

\begin{abstract}
The Australian Astronomical Observatory's `tilting spine' fibre positioning technology has been redeveloped to provide superior performance in a smaller package.  The new design offers demonstrated closed-loop positioning errors of \SI{< 2.8}{\micro\metre} RMS in only five moves (\SI{\sim 10}{\second} excluding metrology overheads) and an improved capacity for open-loop tracking during observations.  Tilt-induced throughput losses have been halved by lengthening spines while maintaining excellent accuracy.  New low-voltage multilayer piezo actuator technology has reduced a spine's peak drive amplitude from \SI{\sim 150}{\volt} to \SI{< 10}{\volt}, simplifying the control electronics design, reducing the system's overall size, and improving modularity.  Every spine is now a truly independent unit with a dedicated drive circuit and no restrictions on the timing or direction of fibre motion.

\end{abstract}


\bstctlcite{IEEE:BSTcontrol}

\section{INTRODUCTION}
\label{sec:intro}

Tilting spine technology, also known as `Echidna' technology, offers simultaneous positioning of hundreds to thousands of densely-packed optical fibres at a telescope's focus.  The technology was originally developed for the Australian Astronomical Observatory's (AAO's) 400-fibre FMOS--Echidna fibre positioner (Subaru telescope) \cite{2000SPIE.4008.1395G, 2003SPIE.4841.1429M} and has since benefited from a number of mechanical refinements to improve positioning performance \cite{2004SPIE.5492..353M, 2008SPIE.7014E..4KM, 2012SPIE.8446E..4WS, 2014SPIE.9151E..1XS}.

\begin{figure}[!p]
    \centering
    \begin{subfigure}[t]{0.55\textwidth}
        \centering
        \includegraphics[width=\textwidth]{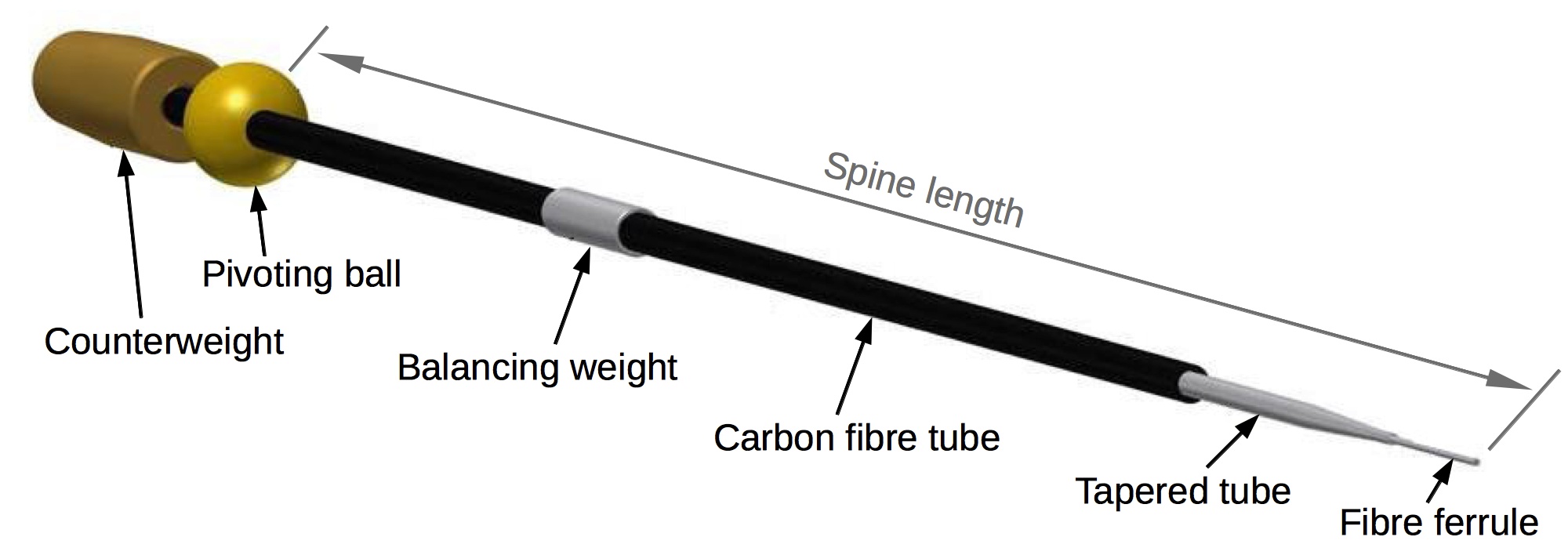}
        \caption{Spine assembly}
        \label{fig:spine_anatomy:a}
    \end{subfigure}%
    ~
    \begin{subfigure}[t]{0.4\textwidth}
        \centering
        \includegraphics[width=\textwidth]{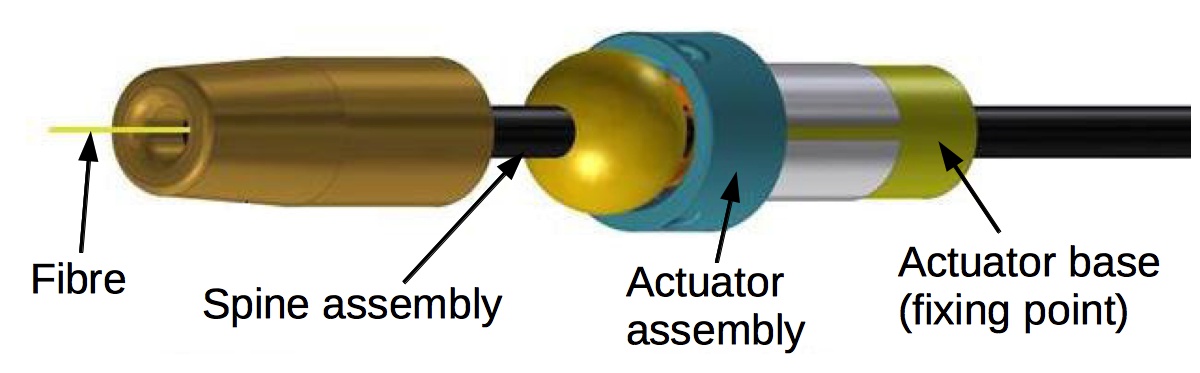}
        \caption{Actuator (motor) assembly}
        \label{fig:spine_anatomy:b}
    \end{subfigure}
    \caption{Spines are passive assemblies (\subref{fig:spine_anatomy:a}) comprising a lightweight rigid tube cemented into a pivoting ball; they are balanced about this ball and therefore operate regardless of the local gravity vector. The fibre passes through the main counterweight, through the tube, and is terminated in a ferrule at the tip of the spine. The specified length of a spine (typically \SI{250}{\milli\metre} max.) is taken from the centre of its ball to its tip. In the existing technology, the spine is held within a cylindrical actuator assembly (\subref{fig:spine_anatomy:b}) by means of a magnetic `cup' mount that is cemented to a piezo tube. The piezo tube is held by a solid mounting structure behind the focal surface of the telescope (see Figure~\ref{fig:spine_packing}). This structure has an oversized clearance hole for the spine, allowing it to tilt by a few degrees in all directions.}
    \label{fig:spine_anatomy}
\end{figure}

\begin{figure} [!p]
\CenterFloatBoxes
\begin{floatrow}
\ffigbox
    {\includegraphics[width=6.5cm]{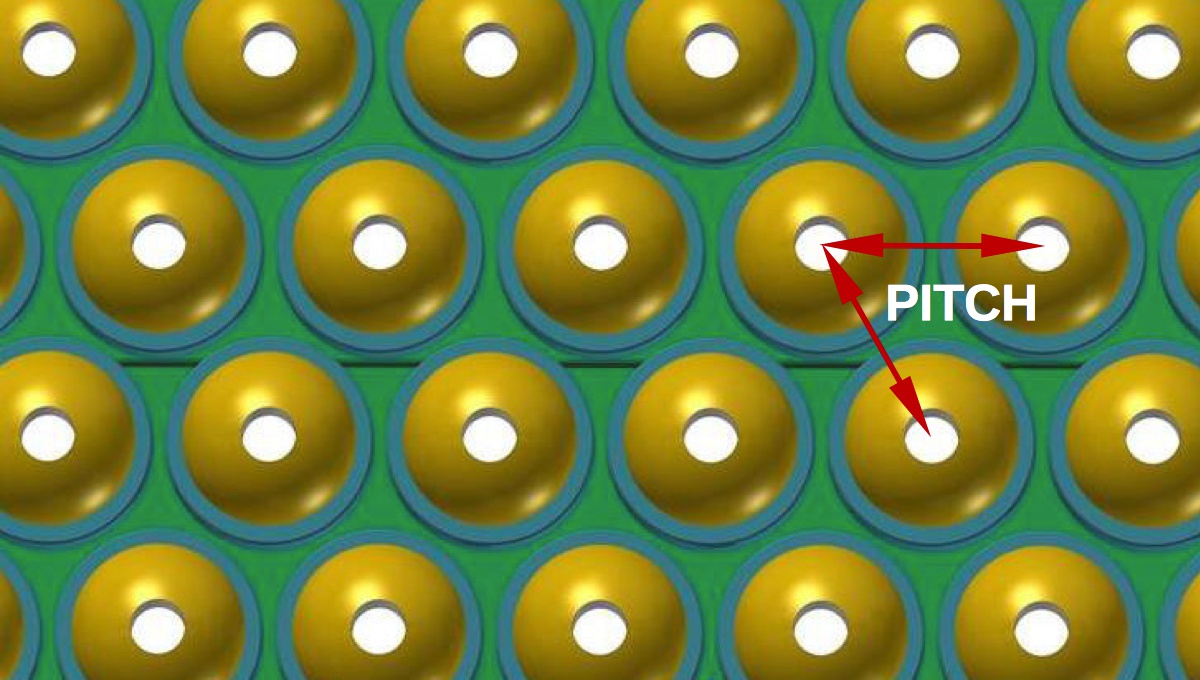}}
    {\caption{Spines are packed in a regular triangular grid. This view is from `behind' the spines, showing only the spine balls (gold) in fixed actuator assemblies (blue).}\label{fig:spine_packing}}
\killfloatstyle
\ttabbox
    {
    \centering
    \small
    \begin{tabular}{p{3cm} p{3cm}}
        \hlinep
        Specification           & Approx.\ value                                    \\
        \hlinep
        Minimum pitch           & \SI{7}{\milli\metre} \cite{2012SPIE.8446E..4WS}   \\
        Patrol radius           & $\num{\sim 1.2}\times$ pitch                      \\
        Min.\ fibre separation  & \SI{< 1}{\milli\metre}                            \\
        Spine length            & \SIrange{140}{250}{\milli\metre}                  \\
        \hlinep
    \end{tabular}
    }
    {\caption{Echidna spines can be closely packed, have a small exclusion radius around their tips, and provide excellent field coverage. It is common for any point on the field to be accessible by at least four and as many as seven different fibres.}\label{tab:spine_stats}}
\end{floatrow}
\end{figure}

The anatomy of a current-generation tilting spine is shown in Figure~\ref{fig:spine_anatomy}.  Each fibre is held in a rigid carbon fibre tube that can freely tilt about a ball at one of its ends.  The tube is balanced about the ball to form a passive spine assembly.  The spine is then inserted through the centre of a cylindrical piezoelectric actuator assembly and magnetically held in place.  Actuator assemblies are mounted on a sold base in a triangular grid (Figure~\ref{fig:spine_packing}).  Table~\ref{tab:spine_stats} shows how fibres can reach far beyond the actuator pitch, giving excellent field coverage.

\begin{figure}[!p]
    \centering
    \includegraphics[width=10cm]{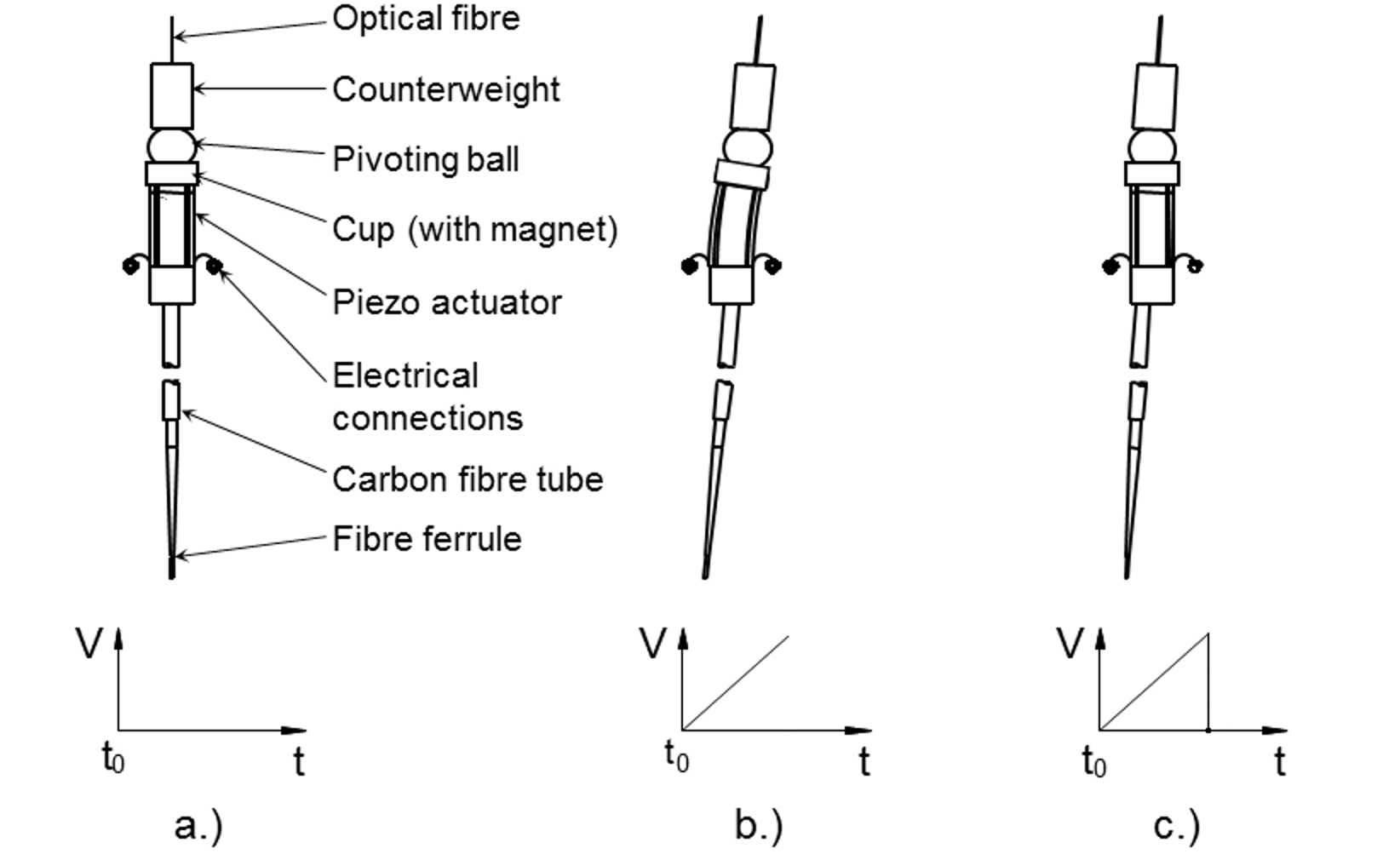}
    \caption{The tilting spine `stick--slip' mechanism is a simple way to achieve discrete displacements of the spine tip using a single piezo tube actuator and a straightforward sawtooth drive waveform: Starting from a neutral position (a), the ramp of the sawtooth steadily bends the actuator and tilts the spine within it (b); the actuator then suddenly returns to its neutral position, leaving the spine behind (c).  The direction of motion ($\pm x, \pm y$) depends on which of the four electrode `quadrants' the drive waveforms are connected to.}
    \label{fig:echidna_principle_stickslip}
\end{figure}

Each spine's actuator assembly can apply small displacements to the surface of the ball, tilting the spine and therefore shifting its tip.  A sawtooth motion profile provides a stick--slip mechanism that moves the fibre in small discrete steps (Figure~\ref{fig:echidna_principle_stickslip}) along one of four orthogonal vectors ($\pm x, \pm y$).  The discrete step size is set by the drive waveform amplitude, the speed of the move is set by the waveform frequency (i.e.\ the stepping frequency), and the move direction depends on which actuator electrodes are driven and at what polarity.  There are generally two types of move: `coarse' moves for when a fibre is far from its target (\SI{\sim 40}{\micro\metre} steps at \SI{\sim 100}{\hertz}); and `fine' moves for when a fibre is near to its target (\SI{\le 10}{\micro\metre} steps at \SI{< 10}{\hertz}).

\begin{figure}[!t]
    \centering
    \begin{subfigure}[t]{\textwidth}
        \centering
        \includegraphics[width=12.4cm]{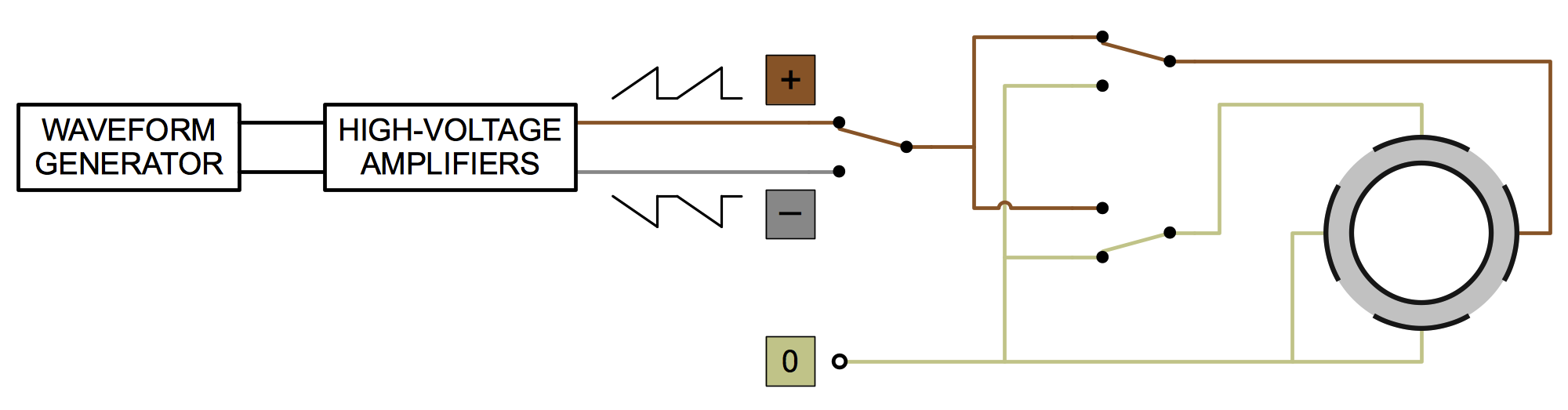}
        \caption{Single spine control circuit}
        \label{fig:echidna_control_electronics:a}
    \end{subfigure}%
    \\
    \smallskip
    \begin{subfigure}[t]{\textwidth}
        \centering
        \includegraphics[width=12.4cm]{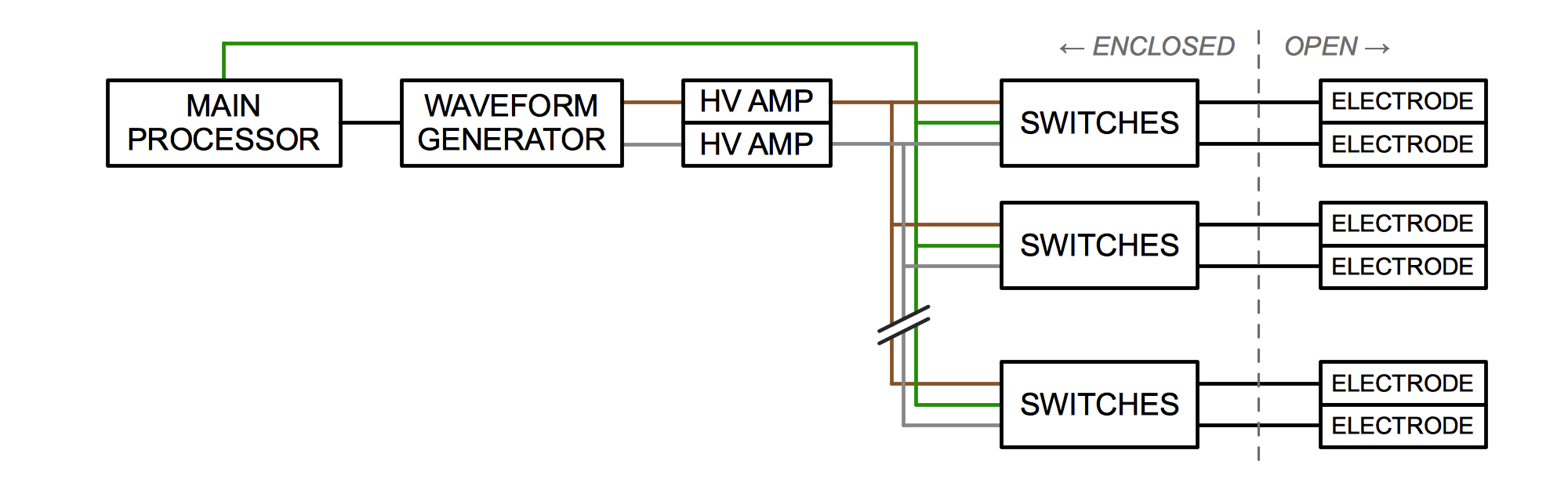}
        \caption{`Shared waveforms' control architecture for multiple spines}
        \label{fig:echidna_control_electronics:b}
    \end{subfigure}
    \caption{The existing tilting spine control architecture routes two `master' waveforms to many spines using an array of switches. Waveforms must be shared in this way because the high-voltage amplifiers required to produce them are large and expensive. Hundreds of spines will often be driven by a single amplifier stage.}
    \label{fig:echidna_control_electronics}
\end{figure}

Traditional piezoceramic actuator technology relies on high electric field strengths to produce appreciable actuator displacements.  This means that Echidna spines have always required drive waveform amplitudes of \SI{\sim 150}{\volt} for normal operation.  Such high voltages present electrical limitations in terms of how spines are controlled, because the necessary amplifiers and power supplies are large and expensive.  This has forced a design compromise whereby many spines share a common pair of waveforms that are routed to actuator electrodes via an array of switches (Figure~\ref{fig:echidna_control_electronics}).

\begin{figure}[!t]
    \centering
    \includegraphics[width=9cm]{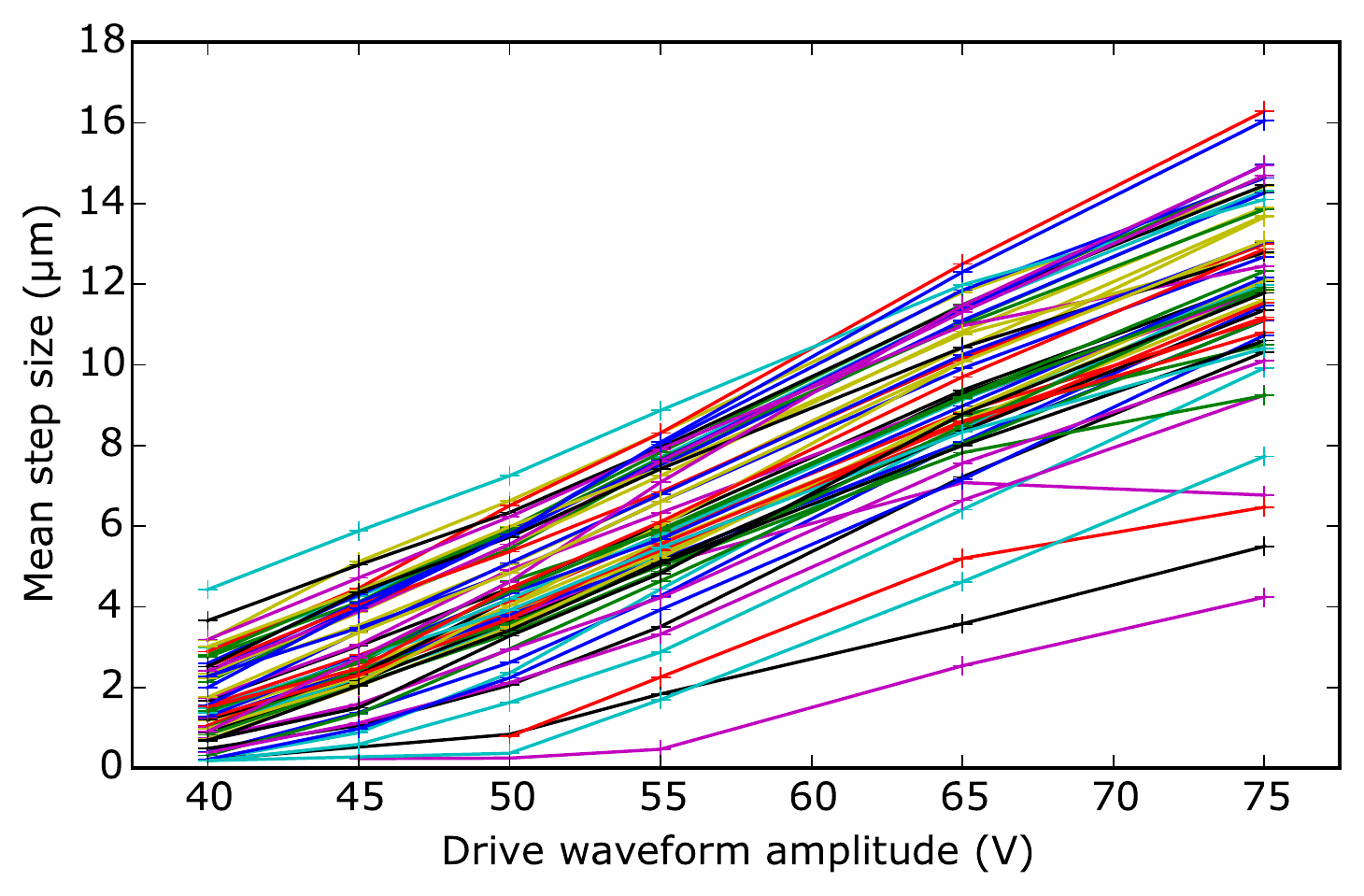}
    \caption{Spine step sizes can vary significantly for the same actuator drive amplitude.  This data was obtained from the AAO's 64-spine AESOP fibre positioner prototype.  Each of the 64 spines is represented by a line.  Note how some reach their lower operational limit at \SIrange{50}{55}{\volt}, which means the global amplitude has to be set higher than this.  This particular prototype system had its fine mode amplitude set at \SI{75}{\volt}, which was sufficient for the relatively loose accuracy requirements of the instrument.  Even so, this amplitude yields inconsistent step sizes, with a range of \SIrange{4}{16}{\micro\metre}.}
    \label{fig:AESOP_step_variation}
\end{figure}

Driving many spines with the same waveforms can seriously limit their performance when very accurate positioning is required.  This is because each spine and actuator pair has its own characteristics that cannot be accounted for on a spine-by-spine basis.  Figure~\ref{fig:AESOP_step_variation} shows the extent of this problem for fine moves, with any one waveform amplitude yielding a wide range of step sizes across spines.  Furthermore, an amplitude must be selected that comfortably exceeds the threshold at which each spine fails to move properly.  It follows that sharing waveforms among many spines results in an overall decrease in accuracy across the entire positioner.

A relevant example is the AAO's Australian ESO Positioner (AESOP) system for the planned 2436-fibre 4MOST instrument (VISTA telescope) \cite{2016SPIE-sheinis-inpress, 2014SPIE.9147E..0MD}.  The current design for this positioner has groups of more than 600 spines sharing the same drive signals, which is likely to result in markedly different positioning resolutions across all fibres.  4MOST's relatively loose positioning accuracy requirement of \SI{\le 10}{\micro\metre} RMS means that this control architecture is acceptable for AESOP.  The problem occurs when tighter accuracy requirements must be met, where achieving these would call for more complex control strategies that increase field reconfiguration time.

A key feature of spine positioners is their high target allocation yield, achievable due to each fibre's large and overlapped patrol area and small exclusion radius around the fibre tip.  The trade-off here is that tilting a fibre away from an incoming beam introduces unavoidable throughput losses due to defocus and focal ratio degradation (FRD) caused by the non-telecentricity.  The relationship between the overall loss and the spine tilt angle is approximately quadratic \cite{2014SPIE.9151E..1XS}, therefore losses are reduced by using longer spines.  Spine length is limited, however, by two factors: i) the spine's minimum achievable step size, which gets larger as spines get longer; and ii) the resonant response of the spine, which becomes more problematic with increased length.  The former point is linked to the compromised accuracy brought about by Echidna's traditional `shared waveforms' control architecture, which is a result of the voltage demands of the actuators.

The goal of this research was to reduce the drive voltage of spine motors, so that: a) spines do not suffer the performance penalty of sharing drive waveforms with others; and b) the control electronics system becomes smaller, leaner, safer, and easier to design.  Through achieving this, the broad aim was to boost the capabilities of the technology by: a) improving the closed-loop positioning performance of the system in terms of accuracy and speed; and b) increasing the maximum spine length limit so that optical losses are reduced.

\section{AN IMPROVED MOTOR DESIGN}
\label{sec:motor}

The existing Echidna actuator assembly has been redesigned to make use of new, low-voltage piezo technology.  This has reduced a spine's drive amplitude from \SI{\sim 150}{\volt} to \SI{< 10}{\volt}.

\begin{figure}[!b]
    \centering
    \begin{subfigure}[t]{0.47\textwidth}
        \centering
        \includegraphics[width=0.9\textwidth]{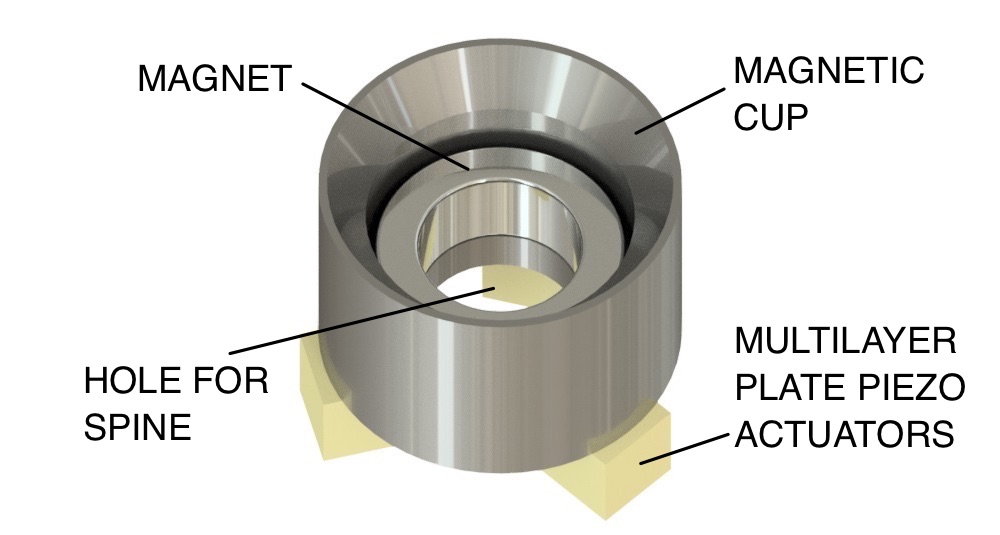}
        \caption{3D model of new motor design}
        \label{fig:motor_design:a}
    \end{subfigure}%
    \begin{subfigure}[t]{0.47\textwidth}
        \centering
        \includegraphics[width=0.9\textwidth]{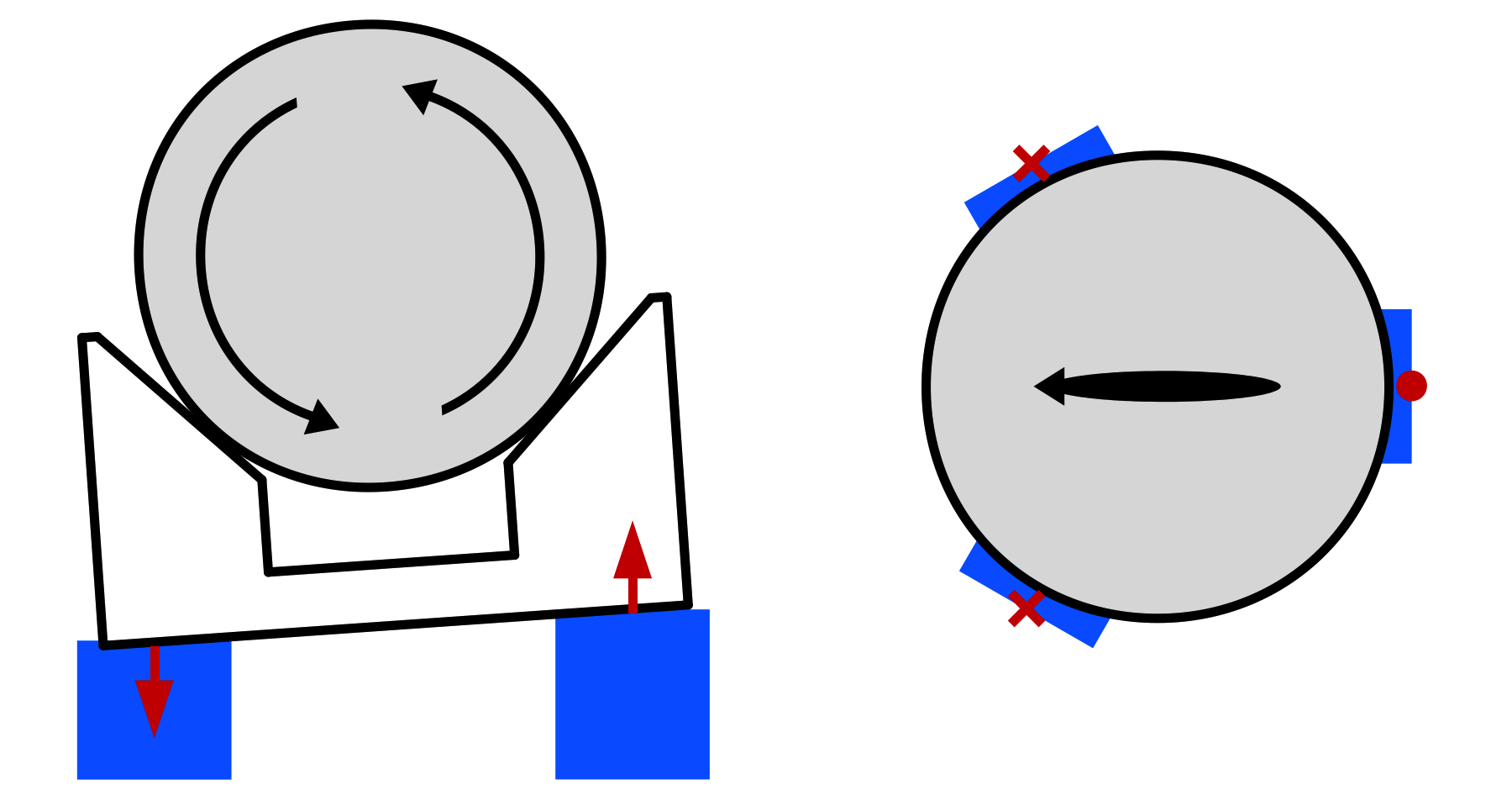}
        \caption{Exaggerated diagram showing actuator forces}
        \label{fig:motor_design:b}
    \end{subfigure}
    \caption{The new motor design has three multilayer piezo actuators fixed underneath the magnetic cup assembly; the actuators act as `pillars' that are able to tilt the cup slightly. An embedded magnet holds the ball in the cup, as in the current Echidna design.  The section view (left in \subref{fig:motor_design:b}) shows two opposing actuators, which is inaccurate but illustrates the principle well.  Spine tubes are not shown. The design shown has a diameter of \SI{8}{\milli\metre}.}
    \label{fig:motor_design}
\end{figure}

Unlike Echidna, which uses a bending tube to shift the magnetic cup assembly sideways under the ball, the new motor design tilts the cup by lengthening or shortening three actuators arranged like pillars (Figure~\ref{fig:motor_design}).  Using actuator displacements parallel to the axis of the spine greatly reduces the coupling of actuator forces with the spine in the orthogonal direction, therefore reducing unwanted resonance.

The new motor's simple rectangular actuators are well suited to modern multilayer piezo manufacturing processes.  Multilayer actuators feature many thin piezoceramic layers with interstitial electrodes, increasing the electric field strength in the material by a factor approximately equal to the number of layers, when compared to an equivalent bulk device.  This means that much lower voltages are required to achieve the same displacements.

The new motor eliminates the operational and design considerations previously made necessary by Echidna's hazardous voltages, such as the need for specialist components, large conductor spacings, bulky connectors, and safety interlock systems to protect personnel.

\begin{figure}[!t]
    \centering
    \includegraphics[width=10cm]{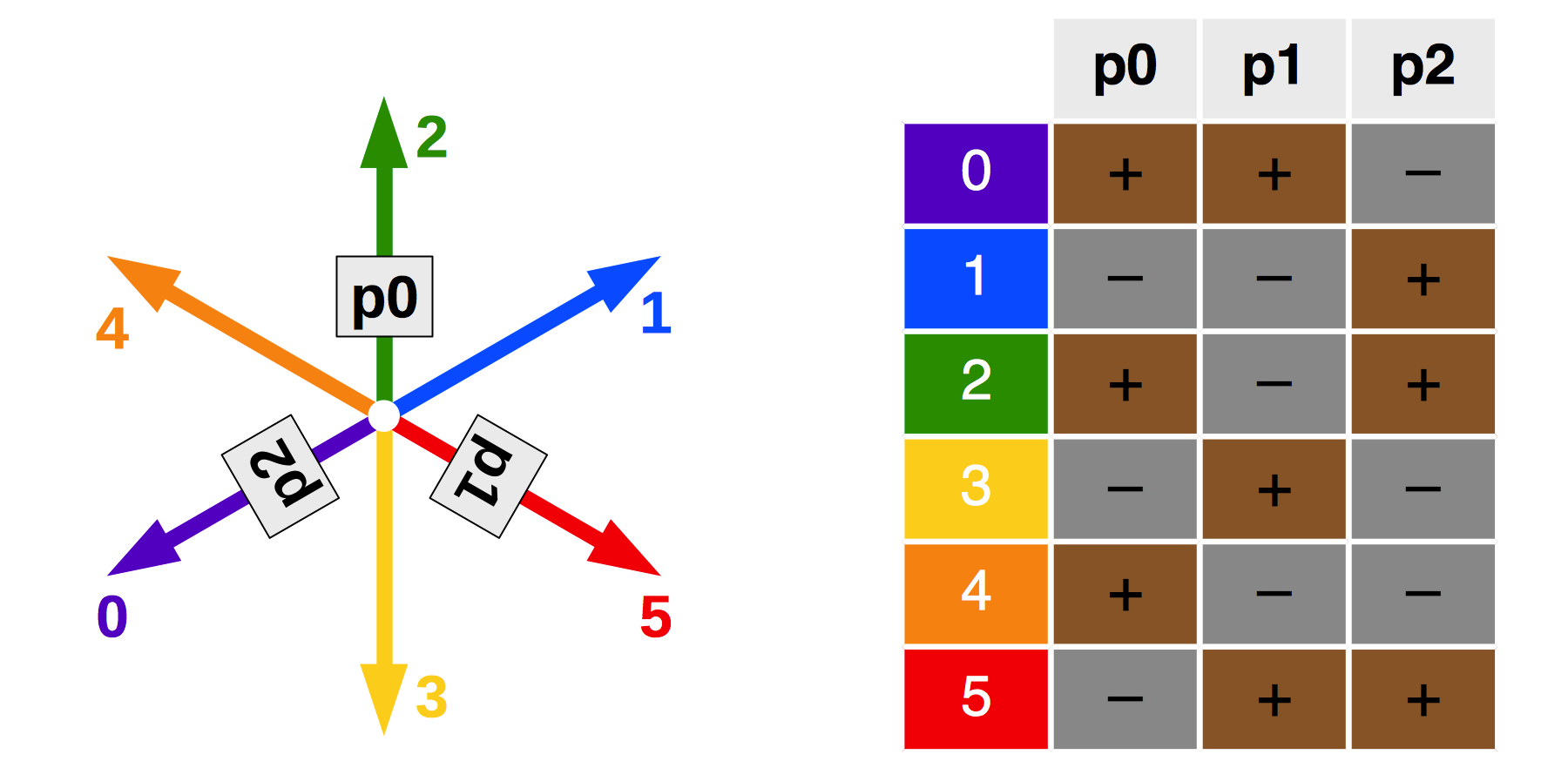}
    \caption{Three actuators (p0, p1, p2) and two waveform polarities ($+$ and $-$) yield spine movement in six possible directions (0--5), depending on the waveform connections. Note that a `negative' waveform is an inverted sawtooth.}
    \label{fig:six_component_dirs}
\end{figure}

Another inherent feature of the motor design is how a spine can be made to move along one of six vectors, rather than the four vectors available previously.  The direction of movement depends on the combination of positive and negative waveforms applied to the actuators (Figure~\ref{fig:six_component_dirs}).

\section{CONTROL ELECTRONICS}
\label{sec:control}

The use of low-voltage piezo actuators has opened up new possibilities in terms of electrical design.  This has improved performance, reduced the size of the control system, and given spines true independence and the ability to move in arbitrary directions.

\begin{figure}[!p]
    \centering
    \begin{subfigure}[t]{0.35\textwidth}
        \centering
        \includegraphics[width=\textwidth]{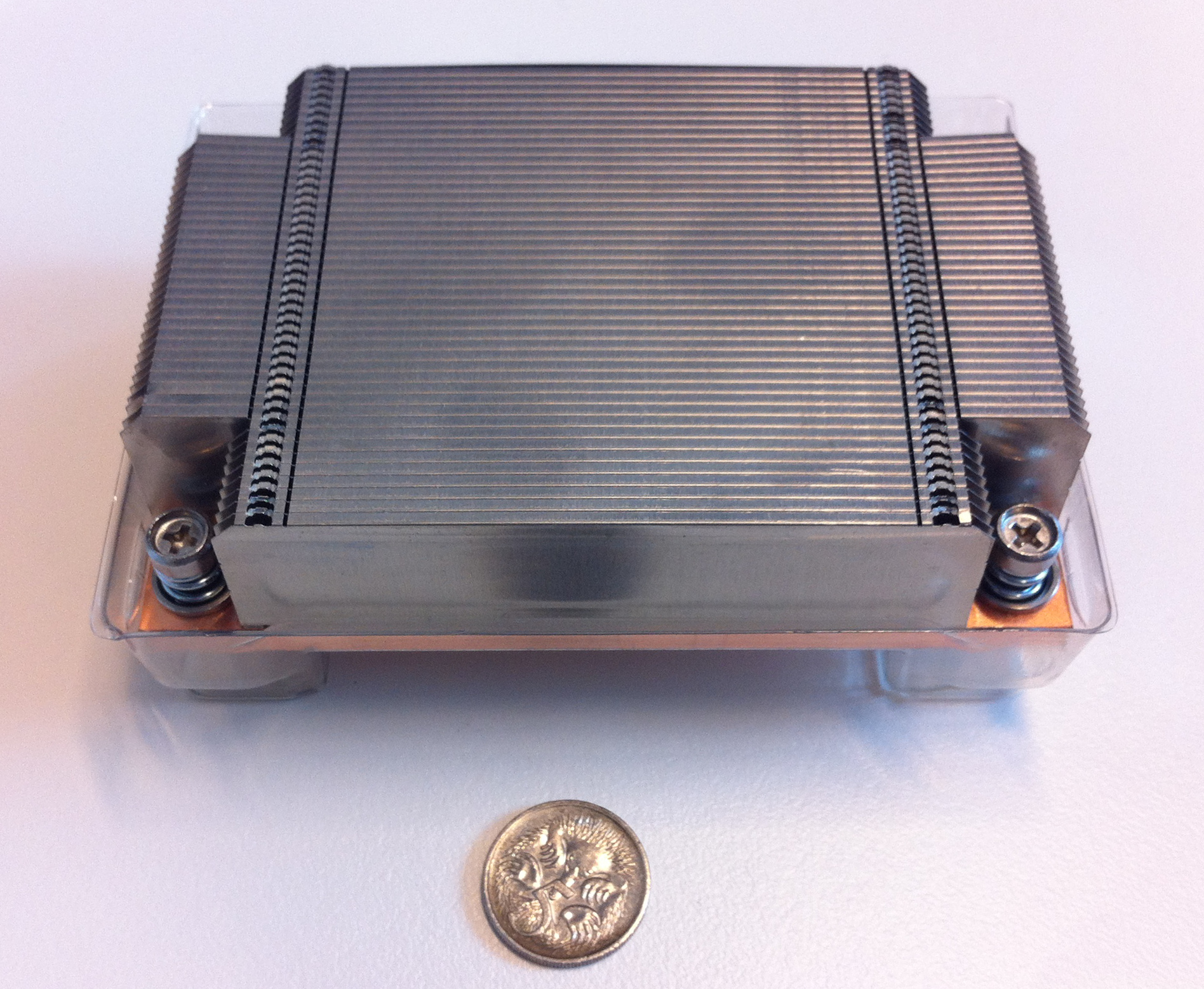}
        \caption{Large \SI{\pm 150}{\volt} amplifier}
        \label{fig:amps_compare:a}
    \end{subfigure}%
    ~
    \begin{subfigure}[t]{0.35\textwidth}
        \centering
        \includegraphics[width=\textwidth]{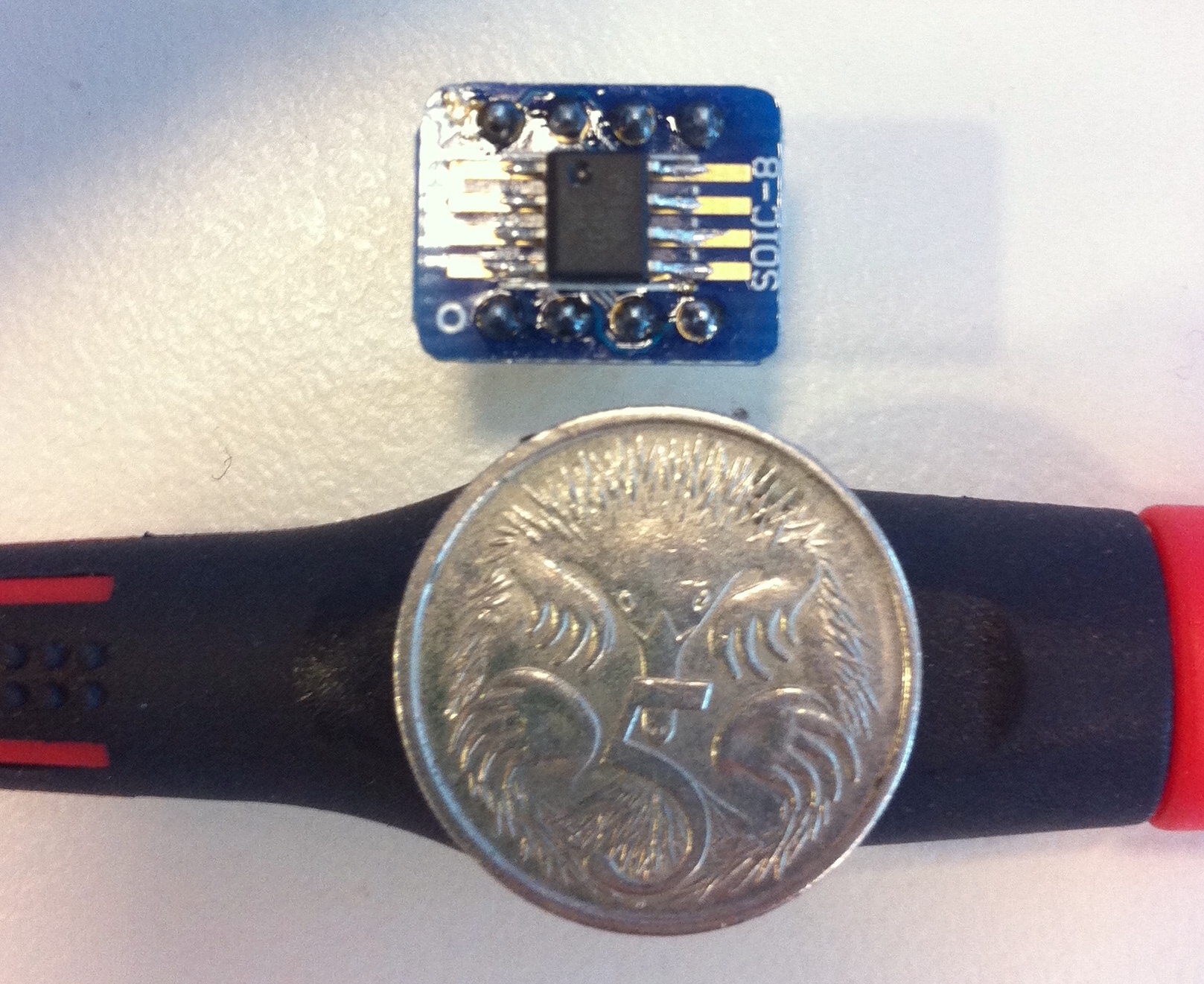}
        \caption{Miniature \SI{\pm 15}{\volt} amplifier}
        \label{fig:amps_compare:b}
    \end{subfigure}
    \caption{The traditional Echidna control electronics has large high-voltage amplifiers (\subref{fig:amps_compare:a}) that can drive over 600 spines with a single pair of drive waveforms, whereas the new motor design allows surface-mount COTS devices (\subref{fig:amps_compare:b}) to be used for individual spines.  The coin shown is an Australian 5~cent piece (diameter \SI{\sim 20}{\milli\metre}).}
    \label{fig:amps_compare}
\end{figure}

\begin{figure}[!p]
    \centering
    \includegraphics[width=11cm]{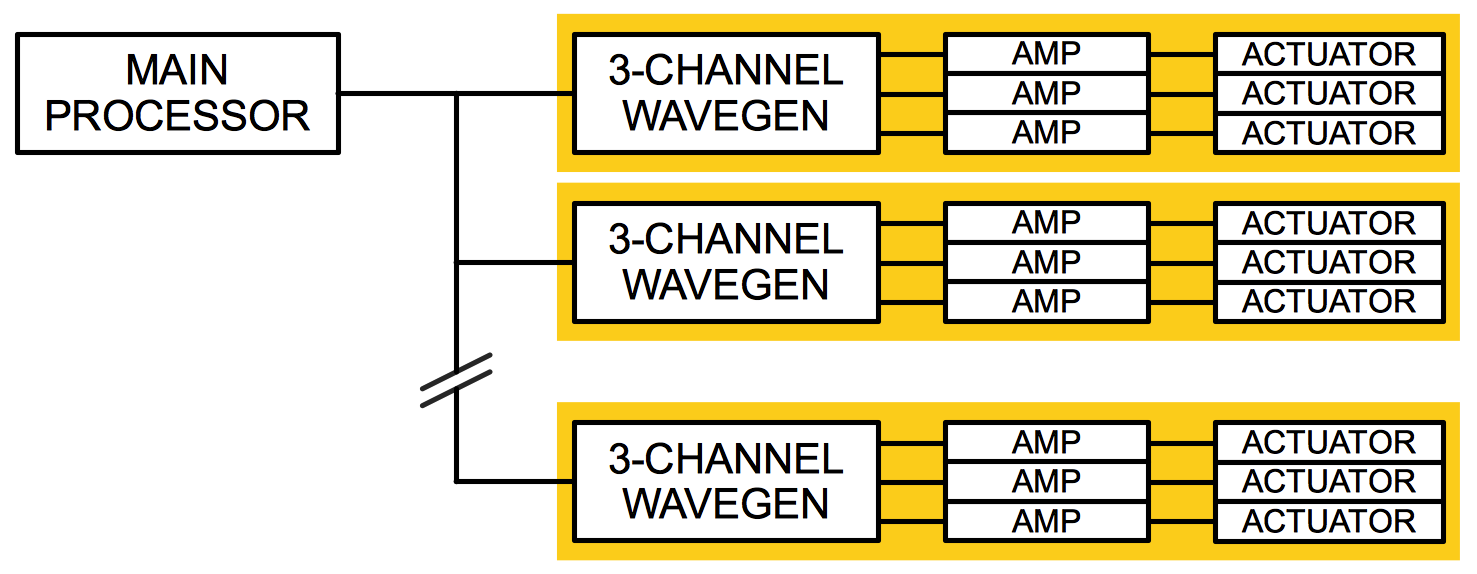}
    \caption{With the new low-voltage motor, it becomes feasible to generate waveforms for every actuator within a motor. This has various performance advantages, but also improves modularity. The level of modularity is highlighted in yellow.}
    \label{fig:elecarch}
\end{figure}

The new motor's modest electrical demands allow the use of small, inexpensive, commercial off-the-shelf (COTS) drive amplifiers where large and expensive ones were used previously (Figure~\ref{fig:amps_compare}).  It follows that, for the same materials cost, every actuator can have its own dedicated drive signals (Figure~\ref{fig:elecarch}).  This eliminates the performance penalty associated with Echidna's traditional `shared waveforms' control architecture, because waveform amplitudes and frequencies can be fine-tuned to the characteristics of individual actuators and spines.  The need for signal routing with switches is removed, and the system becomes more modular.

The existing Echidna technology only allows spines to move in one of four component directions ($\pm x, \pm y$).  This means that fibres will usually undertake a compound move in two of these directions in order to reach a target.  The nature of the shared waveforms architecture is such that all spines must move along the same axis at the same time, before switching to the orthogonal axis to complete the move.  This has two main disadvantages: i) every spine must wait for the longest move to complete before making its next move, increasing positioning time; and ii) collision avoidance often comes with a time penalty, due the limited routes available to each spine.

\begin{figure}[!p]
    \centering
    \begin{subfigure}[t]{0.4\textwidth}
        \centering
        \includegraphics[height=3.5cm]{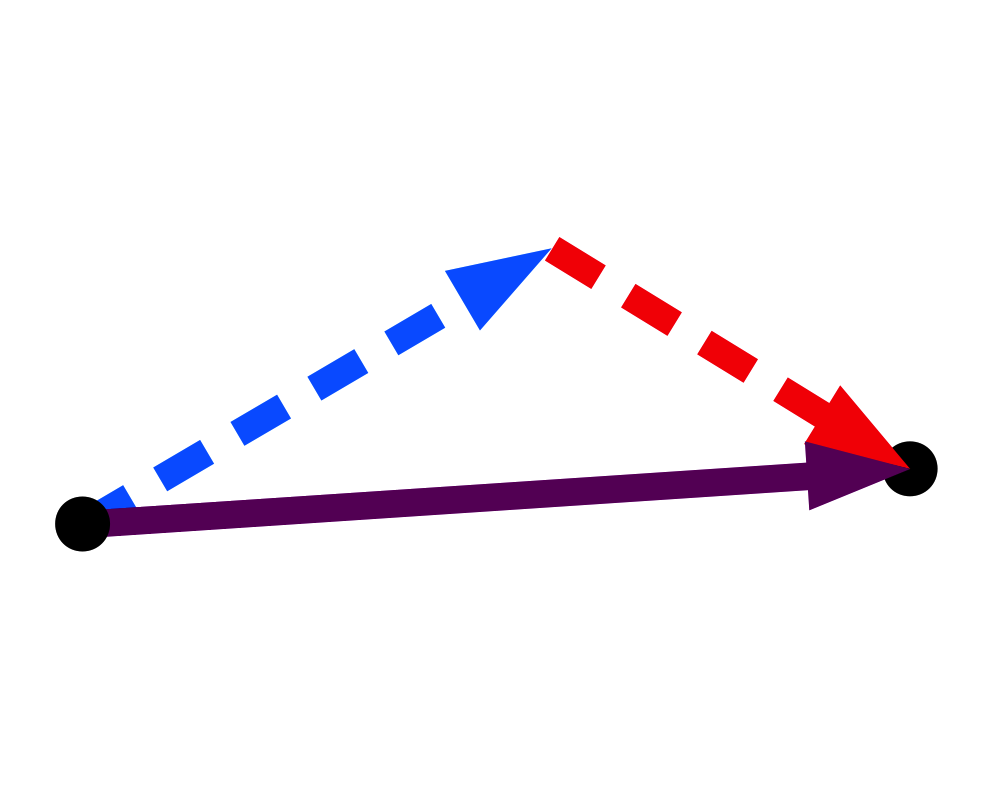}
        \caption{Component vs.\ blended moves}
        \label{fig:blended_moves_principle:a}
    \end{subfigure}%
    \begin{subfigure}[t]{0.4\textwidth}
        \centering
        \includegraphics[height=3.5cm]{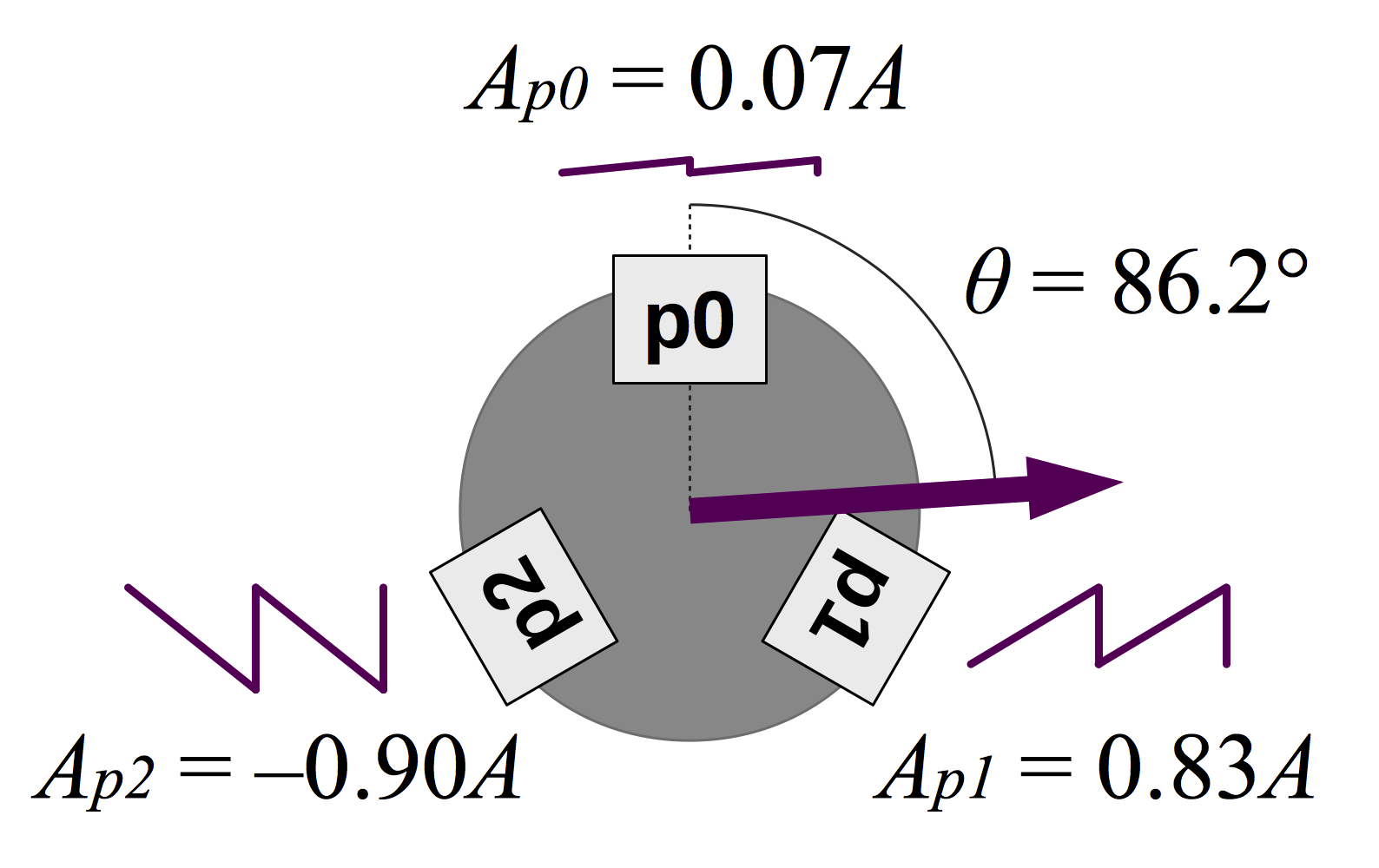}
        \caption{Example waveforms}
        \label{fig:blended_moves_principle:b}
    \end{subfigure}
    \caption{Driving the motor actuators with a particular ratio of amplitudes can make a spine move in any direction, rather than being limited to only six component directions. These have been called `blended' moves.}
    \label{fig:blended_moves_principle}
\end{figure}

The new Echidna motor offers six component move directions instead of four.  This is advantageous in itself, but the new control architecture described above offers something far better, because the ability to set arbitrary drive waveform amplitudes for every actuator in a single motor means that spines can be moved in \emph{any} direction.  This is done by adjusting the ratio of the drive signal amplitudes between each actuator, as illustrated in Figure~\ref{fig:blended_moves_principle}.  The three actuator amplitudes are calculated using simple trigonometry with respect to the actuator positions:

\begin{equation}
    \begin{aligned}
        A_{p0} &= A \cos(000.0 - \theta) \\
        A_{p1} &= A \cos(120.0 - \theta) \\
        A_{p2} &= A \cos(240.0 - \theta)
    \end{aligned}
    \label{eqn:spine_drive_angle}
\end{equation}
where $A$ is the maximum waveform amplitude (i.e.\ smaller $A$ means a smaller step size) and $\theta$ is the desired move direction relative to the first actuator, `p0'.

\begin{figure}[!t]
    \centering
    \begin{subfigure}[t]{0.6\textwidth}
        \centering
        \raisebox{5mm}{\includegraphics[width=0.9\textwidth]{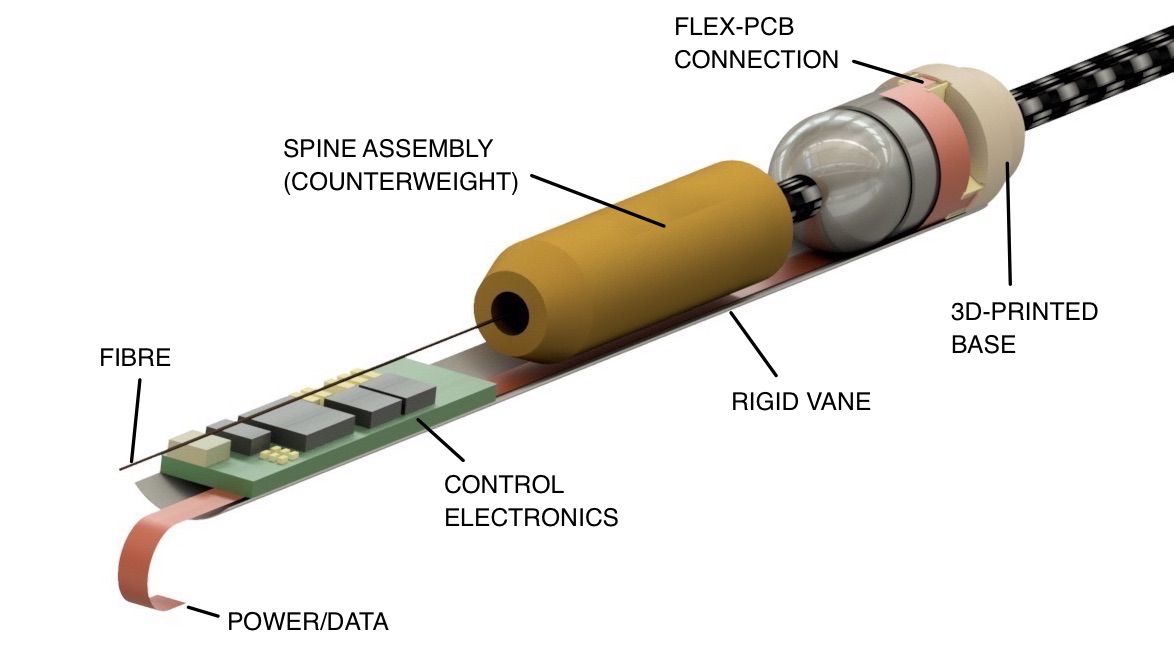}}
        \caption{The stand-alone spine concept}
        \label{fig:stand-alone_assy:a}
    \end{subfigure}%
    \begin{subfigure}[t]{0.26\textwidth}
        \centering
        \includegraphics[width=0.9\textwidth]{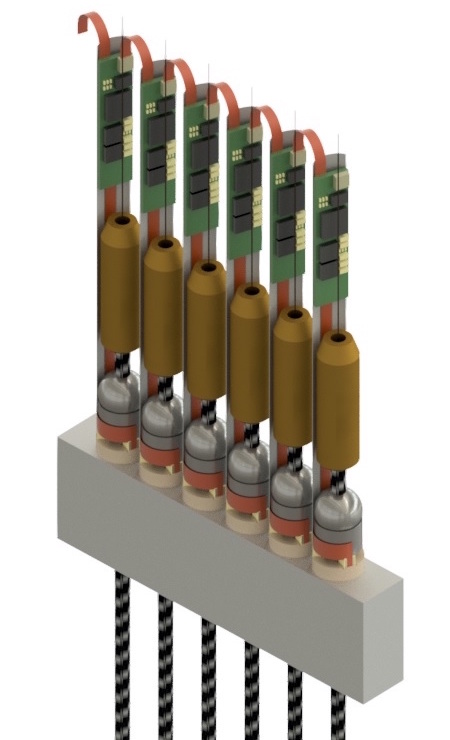}
        \caption{Daisy-chained spines}
        \label{fig:stand-alone_assy:b}
    \end{subfigure}
    \caption{The new Echidna motor ushers in a new level of modularity for tilting spine systems. It is now feasible to produce stand-alone spine assemblies that have a minimal dependence on ancillary electronics systems. This concept design has its control electronics incorporated via a vertical mounting vane, which is cemented to a 3D-printed base. There is also the option of surrounding the entire assembly with a metal tube for enhanced electromagnetic shielding and dust protection. Many spines could populate the science field with daisy-chained flex-PCB connections (\subref{fig:stand-alone_assy:b}). Although difficult to see, there is sufficient clearance around the spine's counterweight for it to tilt freely.}
    \label{fig:stand-alone_assy}
\end{figure}

Another advantage of the new control system design is that the components occupy less circuit board space than in the existing system.  The highly modular architecture also means that the drive circuit for each spine can be treated as a discreet unit and replicated easily for thousands of devices.  This simplifies the design of the control system, which has traditionally been sited around the edge of the science field.  However, the new motor makes it feasible to incorporate drive electronics into every spine assembly.  Figure~\ref{fig:stand-alone_assy} shows a conceptual design for `stand-alone' spines on a \SI{\sim 9}{\milli\metre} pitch.  The integral printed circuit board (PCB) hosts a small microcontroller, a three-channel waveform generator, and miniature power amplifiers.  External connections are simply power (e.g.\ \SI{24}{\volt}) and a standard communications bus (e.g.\ CAN bus).  This removes the need for almost all ancillary electronics around the field and considerably reduces mass.

\begin{table}[!h]
    \caption{Estimated power consumption of a single spine and its control board, using measured and datasheet values.}
    \centering
    \small
    \begin{tabular}{l r}
        \hlinep
        Component                       & Power (RMS)               \\
        \hlinep
        Microcontroller                 & \SI{66}{\milli\watt}      \\
        Amplifiers (quiescent)          & \SI{99}{\milli\watt}      \\
        Actuators (reactive component)  & \SI{34}{\milli\watt}      \\
        Actuators (real component)      & \SI{6}{\milli\watt}       \\
        \hlinep
        Total                           & \SI{205}{\milli\watt}     \\
        \hlinep
    \end{tabular}
    \label{tab:mlp_control_ckt_power}
\end{table}

The power demand of the new motor during positioning was measured in the lab and is summarised in Table~\ref{tab:mlp_control_ckt_power} alongside datasheet values for the other major components in the conceptual miniaturised control circuit.  From this we obtain a total estimated power of \SI{\sim 0.2}{\watt} for a motor running continuously in a coarse-positioning mode.  If we assume a positioner with 2500 spines and a conservative duty cycle estimate of \SI{10}{\percent}, then the average power dissipated into the positioner's surroundings is \SI{51}{\watt} with exposed electronics (stand-alone spines).  If all control electronics were housed in temperature-controlled enclosures around the edge of the field, as is the usual format, then the average power dissipation of all actuators is just \SI{1.5}{\watt}.

\section{PERFORMANCE TESTING}
\label{sec:results}

\begin{figure}[!p]
    \centering
    \begin{subfigure}[t]{0.28\textwidth}
        \centering
        \includegraphics[width=0.85\textwidth]{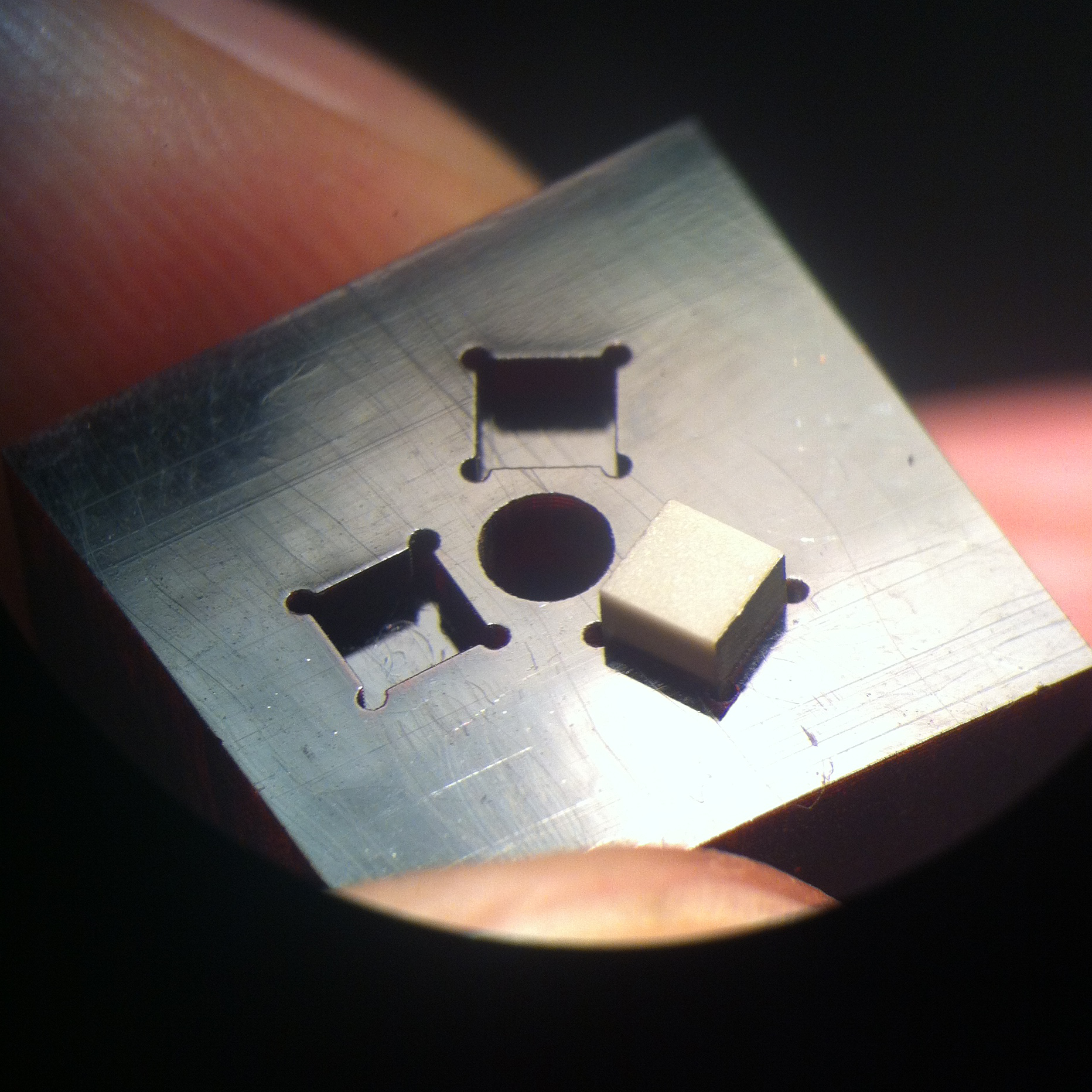}
        \caption{Piezo alignment jig}
        \label{fig:mlp_proto_1:a}
    \end{subfigure}%
    \begin{subfigure}[t]{0.28\textwidth}
        \centering
        \includegraphics[width=0.85\textwidth]{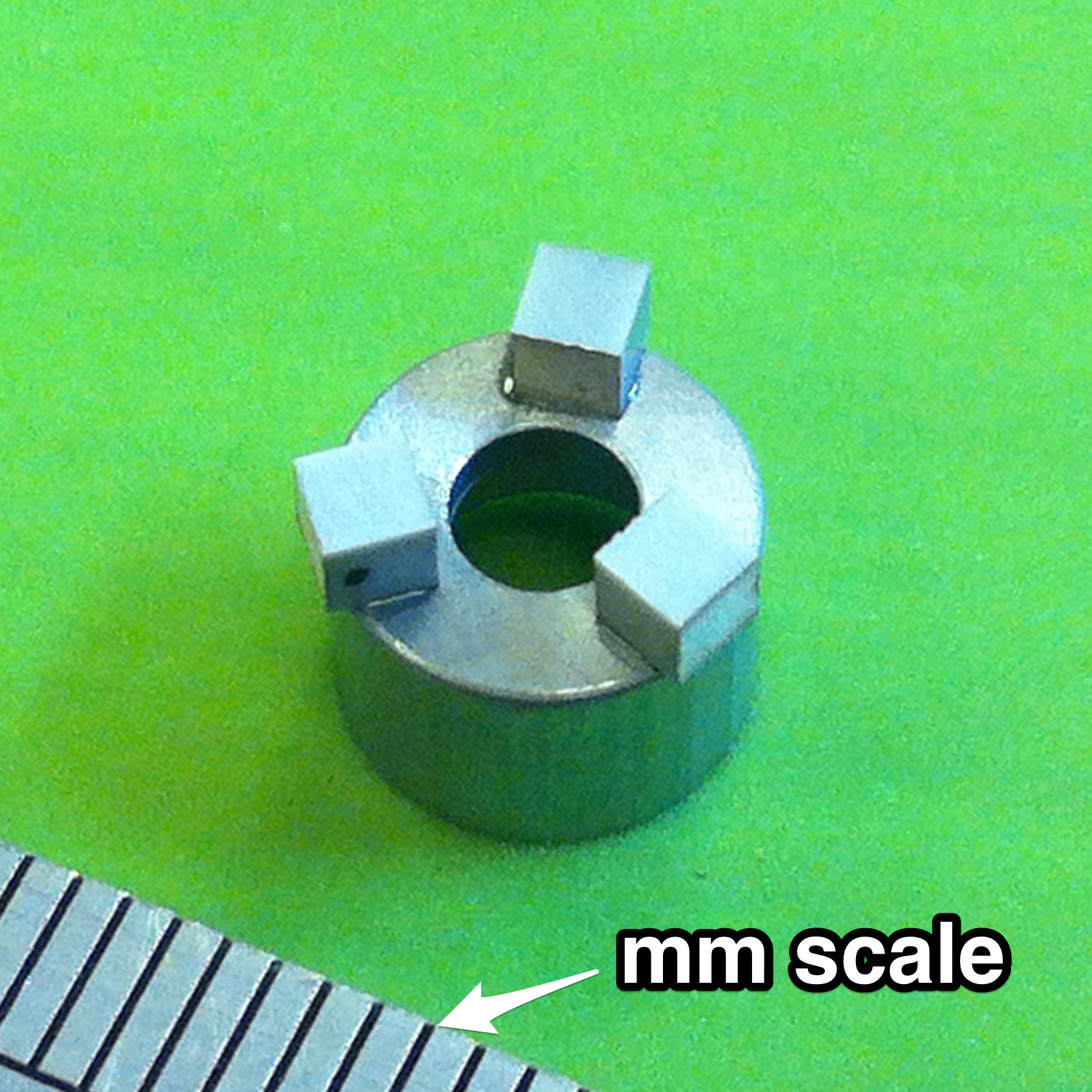}
        \caption{Piezos under cup}
        \label{fig:mlp_proto_1:b}
    \end{subfigure}%
    \begin{subfigure}[t]{0.28\textwidth}
        \centering
        \includegraphics[width=0.85\textwidth]{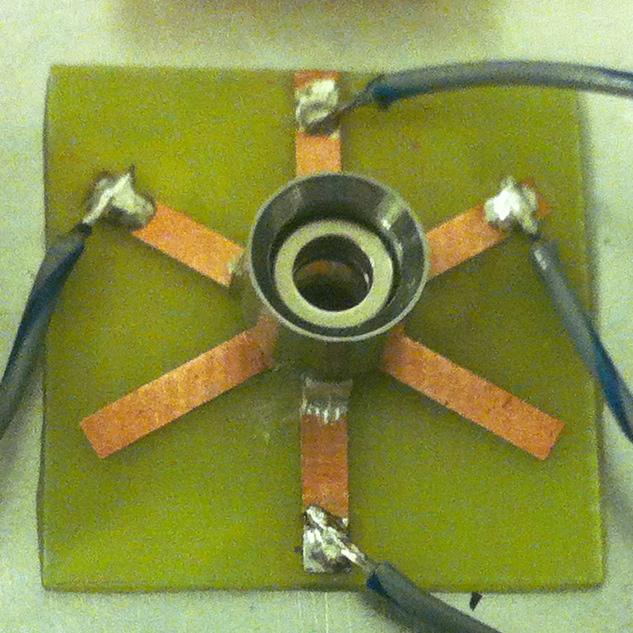}
        \caption{Motor on test PCB}
        \label{fig:mlp_proto_1:c}
    \end{subfigure}
    \caption{Several prototype motors have been built. This prototype has a breakout PCB for the purpose of testing, where ordinarily the electrical connections would exit the PCB around the edge of the science field (unless on-board electronics are used). The motor only requires three electrical connections, despite four being shown.}
    \label{fig:mlp_proto_1}
\end{figure}

Long-term laboratory testing of several new motor prototypes (Figure~\ref{fig:mlp_proto_1}) on a purpose-built test rig has revealed excellent performance and longevity, and has yielded a significant increase to the maximum spine length limit.

\subsection{Closed-loop positioning}

Spine positioning is a closed-loop iterative process: A metrology camera images the entire field and measures the positions of all fibres to a high accuracy.  The fibres are back-illuminated during this process.  Every spine that is not within a given distance from its target position is commanded to move according to its calibration data and any necessary collision avoidance strategies.  With the new motor design, moves can be in any direction and are completely independent of other spines, although it is usually preferable to wait for all spines to finish moving before taking another measurement image due to metrology dominating the total reconfiguration time.  This process is repeated until all spines are close enough to their target, or after a maximum number of iterations.

\begin{figure}[!p]
    \centering
    \begin{subfigure}[t]{0.5\textwidth}
        \centering
        \includegraphics[width=\textwidth]{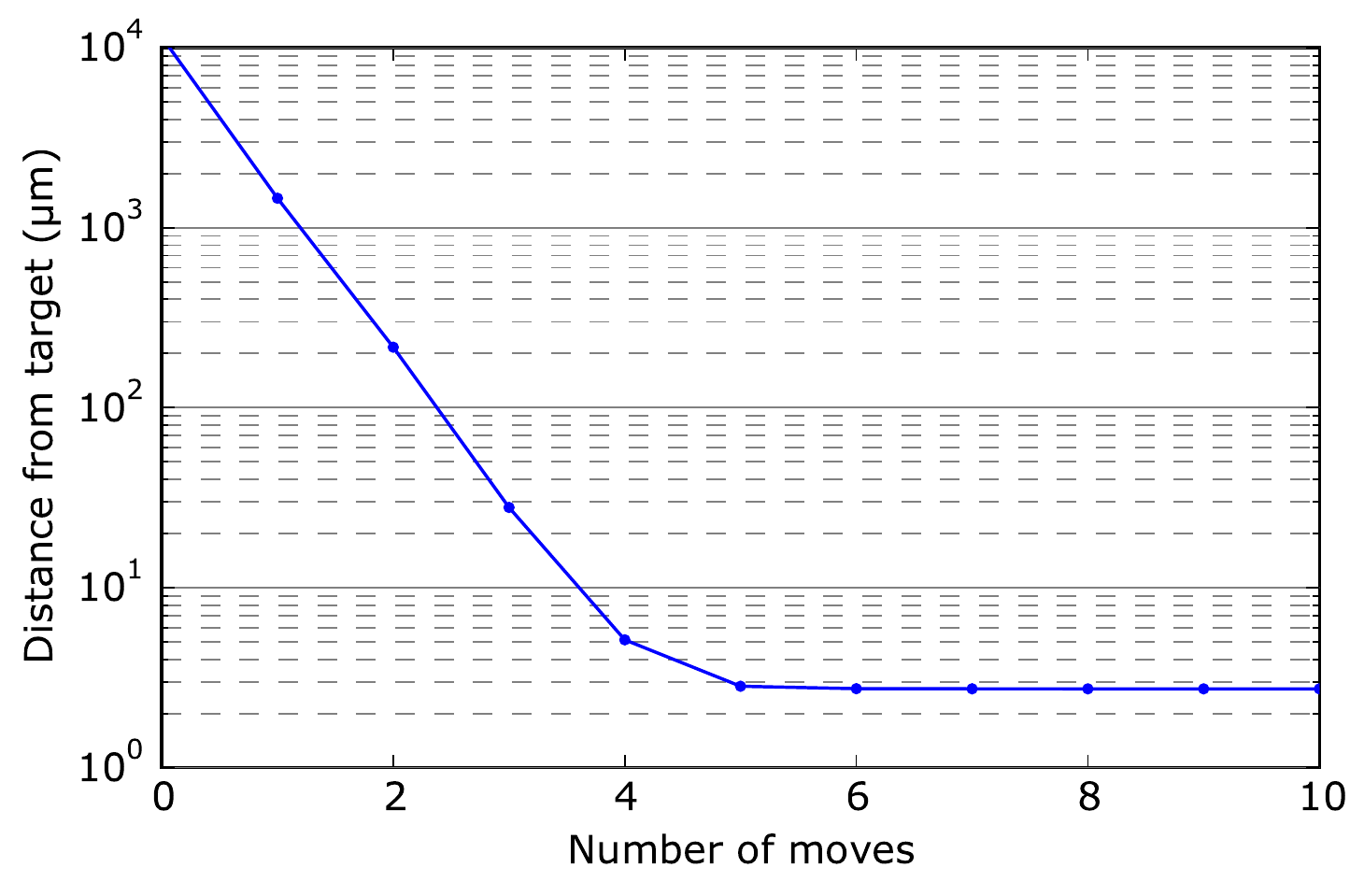}
        \caption{RMS positioning error}
        \label{fig:longevity_proto5_multi:a}
    \end{subfigure}%
    \begin{subfigure}[t]{0.5\textwidth}
        \centering
        \includegraphics[width=\textwidth]{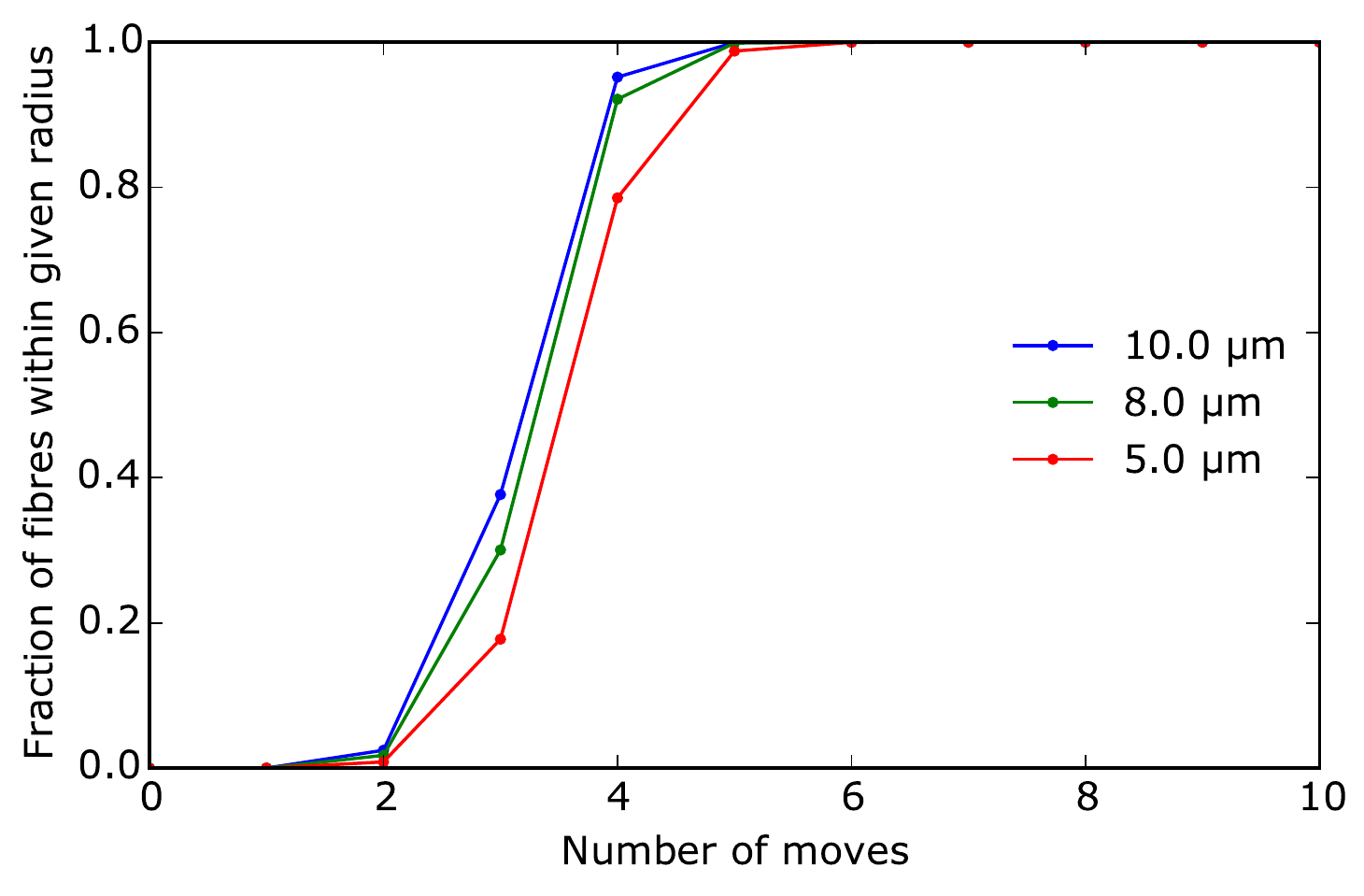}
        \caption{Cumulative convergence}
        \label{fig:longevity_proto5_multi:b}
    \end{subfigure}
    \\
    \smallskip
    \begin{subfigure}[t]{0.5\textwidth}
        \centering
        \includegraphics[width=\textwidth]{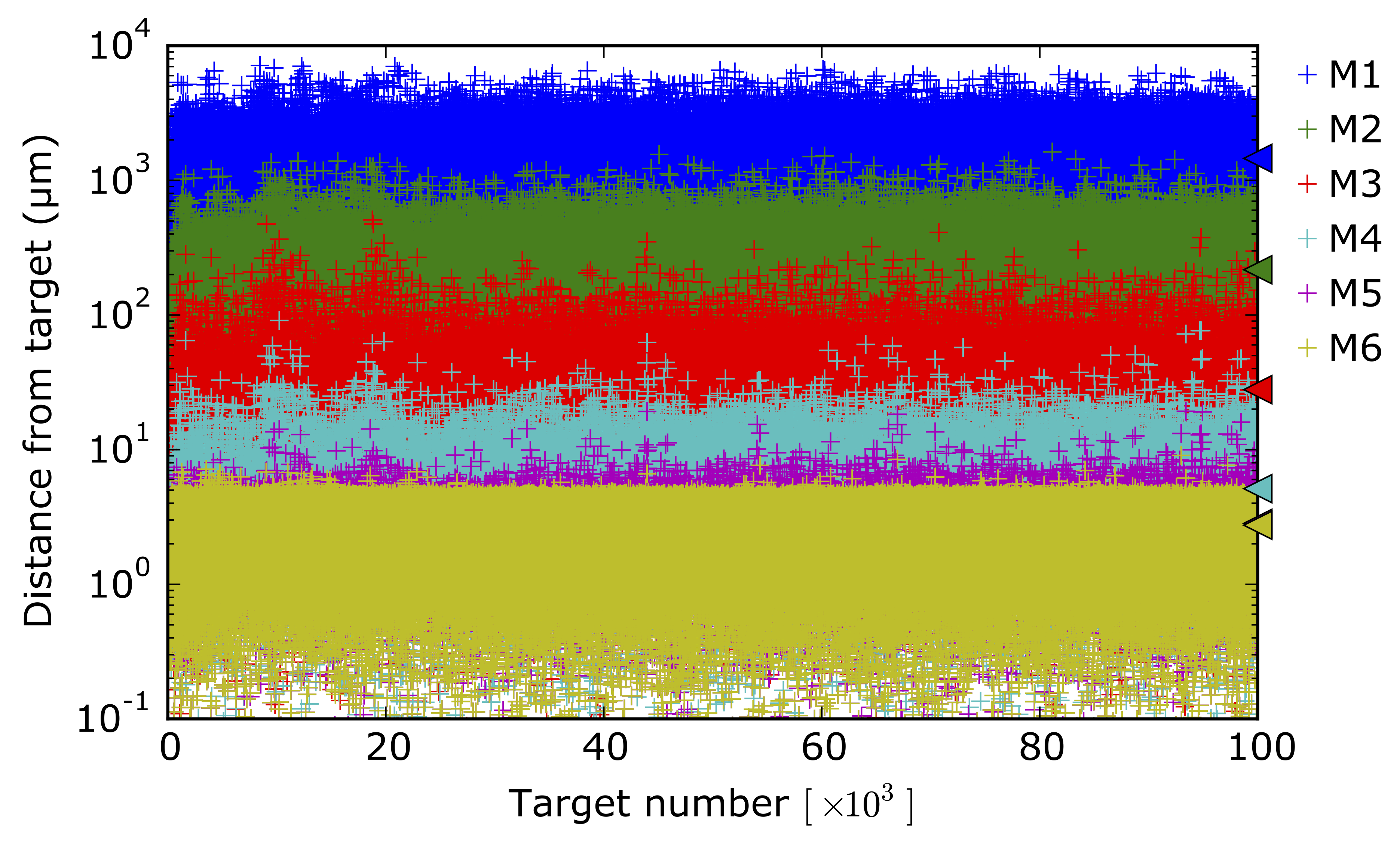}
        \caption{Accuracy over time}
        \label{fig:longevity_proto5_multi:c}
    \end{subfigure}%
    \begin{subfigure}[t]{0.5\textwidth}
        \centering
        \includegraphics[width=\textwidth]{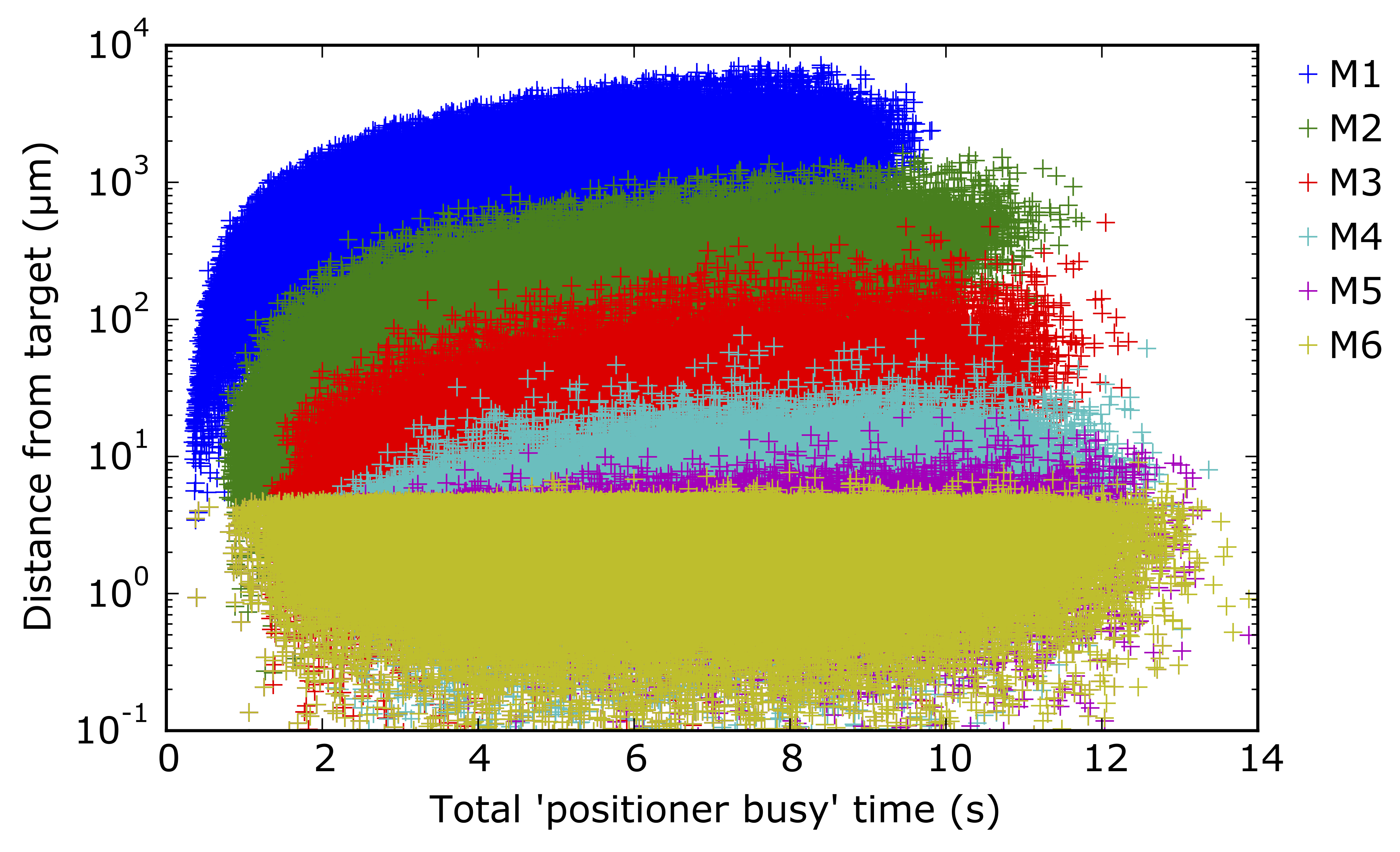}
        \caption{Total move time}
        \label{fig:longevity_proto5_multi:d}
    \end{subfigure}
    \caption{Throughout \num{100000} successive targets (a period of \SI{\sim 60}{\day}) the prototype motor showed consistent and accurate closed-loop performance with an RMS positioning error of \SI{< 2.8}{\micro\metre} after five moves, with negligible improvement thereafter.  The maximum final error for across all targets was \SI{5.8}{\micro\metre}. M1--M6 in (\subref{fig:longevity_proto5_multi:c}) and (\subref{fig:longevity_proto5_multi:d}) are move numbers. Details: \SI{250}{\milli\metre} spine; \SI{9.0}{\volt}/\SI{84}{\hertz} coarse; \SI{3.5}{\volt}/\SI{5}{\hertz} fine; $R_{patrol}\! =\! \SI{11.5}{\milli\metre}$; $Zd\! =\! \SI{45}{\degree}$; blended moves with adaptive calibration.}
    \label{fig:longevity_proto5_multi}
\end{figure}

Figure \ref{fig:longevity_proto5_multi} shows the closed-loop performance of a single prototype motor in a continuous positioning test of \num{100000} random targets across a circular patrol area of radius \SI{11.5}{\milli\metre}.  We see that the spine under test achieves \SI{< 2.8}{\micro\metre} errors in five moves, at which point \SI{98.76}{\percent} of cycles are within \SI{5.0}{\micro\metre} of the target.  \SI{99.97}{\percent} of cycles are within \SI{5.0}{\micro\metre} in six moves, and the maximum final error across all targets is \SI{5.8}{\micro\metre}.

We also see that a maximum of \SI{14}{\second} is spent waiting for moves to complete.  If we assume a \SI{5}{\second} metrology overhead per move and a \SI{1}{\second} control overhead (routing, communications) per move, then the total field reconfiguration time would be \SI{\sim 44}{\second} for \SI{< 2.8}{\micro\metre} RMS accuracy.

To date, three prototype motors have been tested for closed-loop positioning and all have demonstrated \SI{< 2.8}{\micro\metre} RMS accuracy under the same test conditions.

\subsection{Open-loop repeatability}

Good open-loop repeatability is advantageous for two reasons: i) full closed-loop positioning will be faster, requiring fewer corrective moves; and ii) objects that move due to differential refraction and/or field rotation can be tracked across the focal surface without disturbing on ongoing observation.

Early tests of open-loop repeatability for the motor focussed on the stability of step sizes across many successive calibration routines, often revealing very little scatter in the data.  This paints a misleading picture, however, because in reality there are more variables to consider.

\begin{figure}[!h]
    \centering
    \includegraphics[width=9cm]{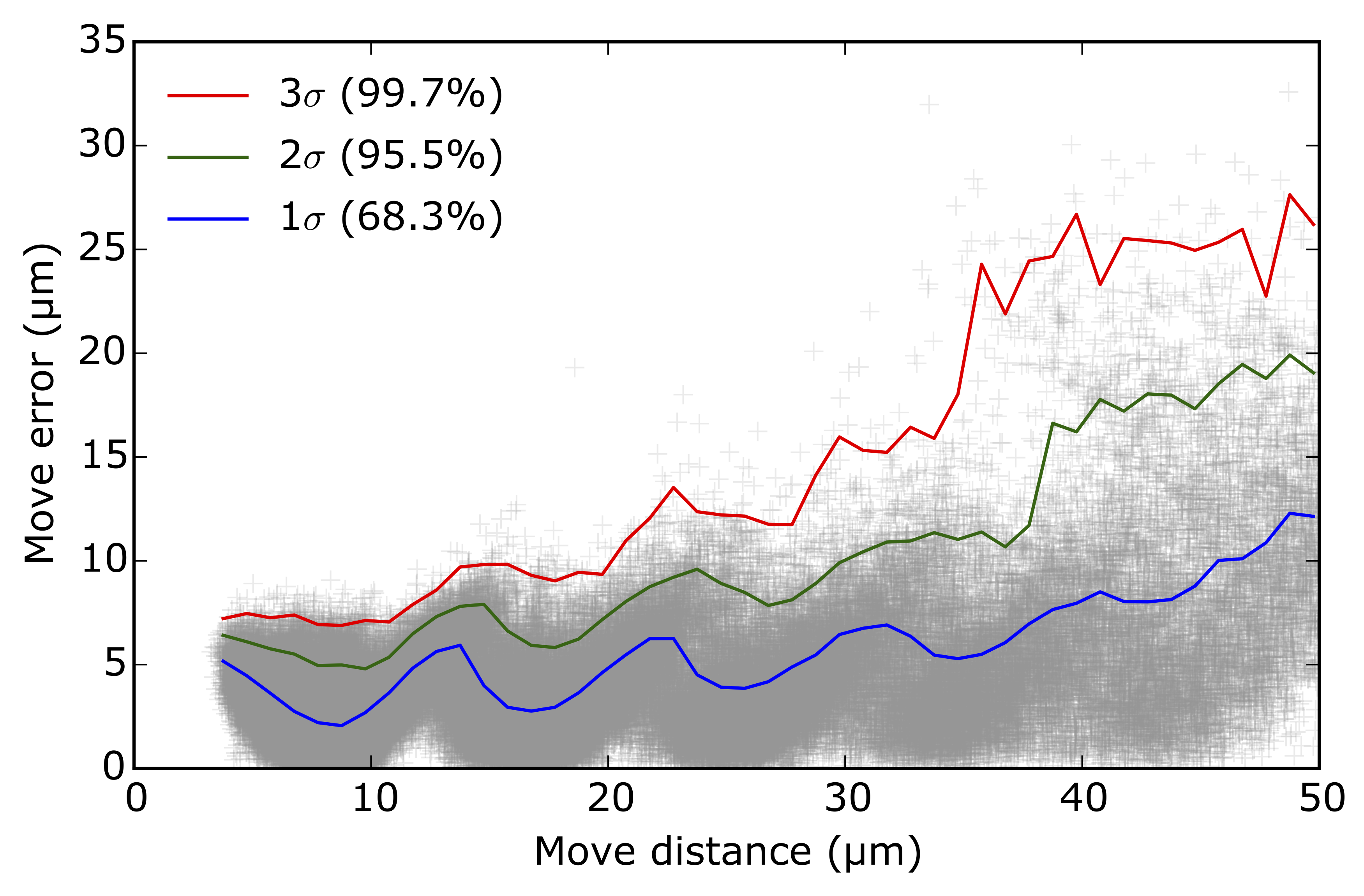}
    \caption{The open-loop repeatability of any single move depends on the distance of the move. These plots encompass errors in direction as well as magnitude; they show the distance between the actual result of a move and the expected result. \num{110000} moves are shown, taken from a long-term closed-loop positioning test. Only short moves are plotted as these are most significant for object tracking. Note the periodic evidence of step multiples, and the population of coarse mode moves beginning to appear at a distance of \SI{\sim 40}{\micro\metre}.}
    \label{fig:spine_testing_openloop}
\end{figure}

Figure~\ref{fig:spine_testing_openloop} reveals the true ability of the tested prototype to move from a random point to another random point, based on the distance of the move.  We see much larger relative errors for the finest moves.  Larger distances, made up of multiple steps, have significantly smaller relative errors.  A possible explanation for this is hysteresis in the piezo actuators; it is believed that a low-order model can improve single-step performance in the future.  This is a known phenomenon from the existing Echidna technology, although the open-loop performance of the new motor is slightly better overall.

\subsection{Increasing spine length}

A key aim of this work was to improve the motor design to a degree that would allow spines to be lengthened, reducing the optical losses caused by tilt-induced FRD.

Spine length has previously been limited to \SI{\sim 250}{\milli\metre} by the minimum achievable step size of the spines and by mechanical limitations (resonance).  The new motor was designed to overcome the latter limitation due to its reduction of forces orthogonal to the spine (i.e.\ actuators now move along the spine axis).

\begin{figure}[!h]
    \centering
    \begin{subfigure}[t]{0.5\textwidth}
        \centering
        \includegraphics[width=\textwidth]{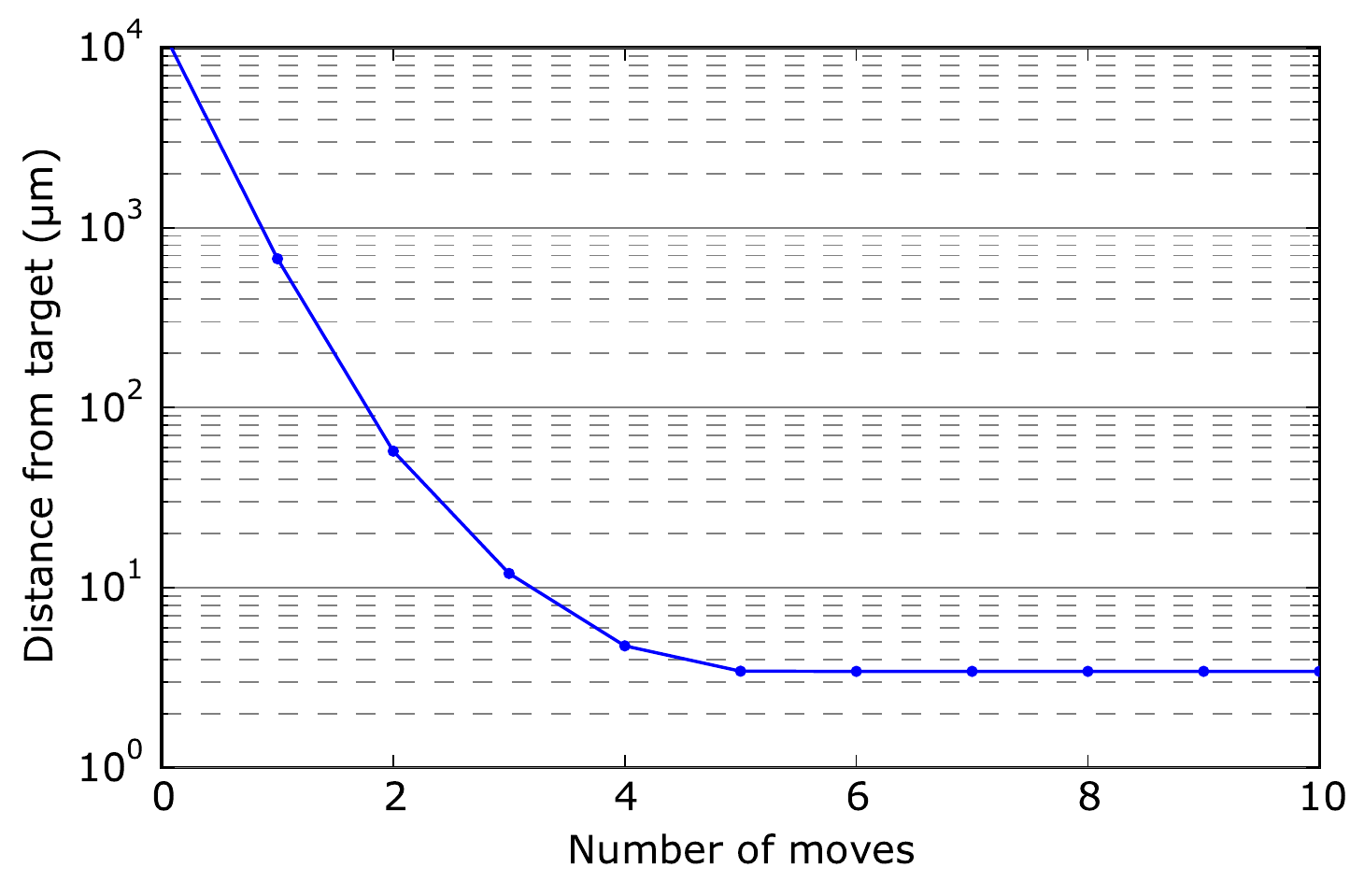}
        \caption{\SI{309}{\milli\metre} spine (\SI{-35}{\percent} optical loss)}
        \label{fig:spine_testing_length:a}
    \end{subfigure}%
    \begin{subfigure}[t]{0.5\textwidth}
        \centering
        \includegraphics[width=\textwidth]{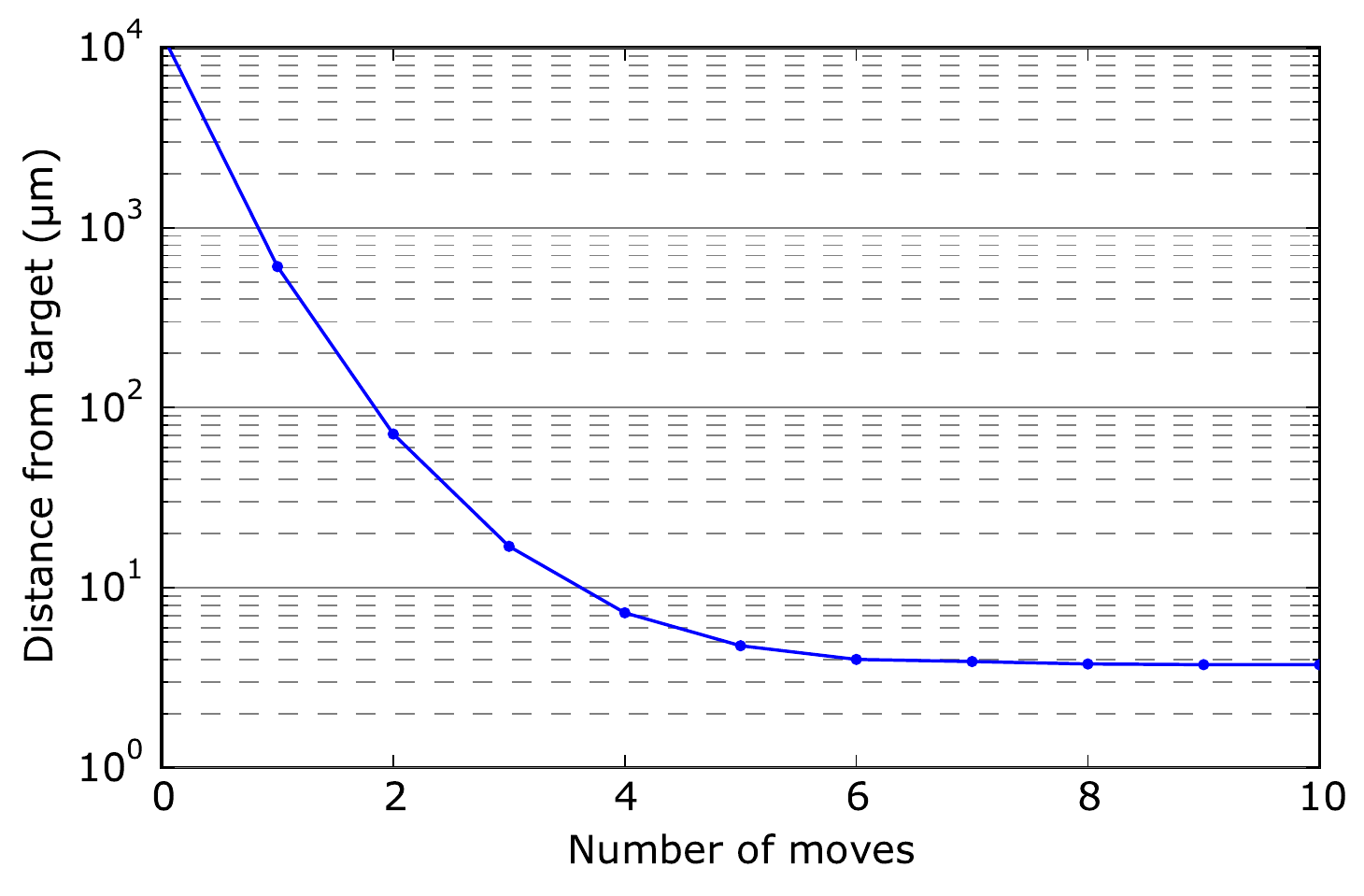}
        \caption{\SI{354}{\milli\metre} spine (\SI{-50}{\percent} optical loss)}
        \label{fig:spine_testing_length:b}
    \end{subfigure}
    \caption{Longer spines performed very well in two short positioning tests of \num{200} targets. The test setup was otherwise equivalent to that of Figure~\ref{fig:longevity_proto5_multi:a}. The positioning errors after each move scale very closely with spine length, as hoped.}
    \label{fig:spine_testing_length}
\end{figure}

The relationship between tilt angle (and therefore spine length) and optical losses is approximately quadratic \cite{2014SPIE.9151E..1XS}.  With this in mind, two experiments were carried out with spine length increases of $\num{\sim 1.2}\times$ and $\num{\sim 1.4}\times$, corresponding to optical loss reductions of \SI{\sim 35}{\percent} and \SI{\sim 50}{\percent} respectively.  Figure~\ref{fig:spine_testing_length} shows successful positioning results in both cases, with accuracies reducing only by the ratio of length increase.  The longest spine is able to achieve \SI{4.0}{\micro\metre} RMS accuracy in six moves.

This is further proof that the new motor is superior to the existing technology; optical losses can be halved while maintaining excellent positioning performance.

\bigskip

\subsection{Longevity}

The final test of the motor conducted for this research concerned the mechanical robustness of the design and its ability to maintain good performance over a very large number of positioning cycles.

The lifetime of an instrument is highly dependent on the observing strategy.  The current 4MOST project is a good example of high demands on a positioner throughout its planned \SI{15}{\year} operational life, because of its relatively short `observation block' size (exposure time) of \SI{20}{\minute}.  At 350 operational nights per year, the spines in 4MOST's AESOP positioner must be able to complete at least \num{157500} closed-loop positioning cycles.

At the time of writing, one prototype of the new motor design has completed \num{> 250000} closed-loop positioning cycles in total, with no signs of performance degradation (\SI{< 2.8}{\micro\metre} RMS, \SI{5.8}{\micro\metre} max.).  This has also been a test of the updated control electronics design and operation mode (blended moves).  A zenith angle of \SI{45}{\degree} was initially chosen, but has been changed to \SI{0}{\degree} (spine pointing up), \SI{90}{\degree} and \SI{180}{\degree} (spine pointing down) with no measurable effect on performance.

\section{CONCLUSIONS AND PROSPECTS}
\label{sec:conclusions}

A new Echidna motor, based on low-voltage multilayer piezoelectric technology, has been designed and prototyped.  The design stays true to the original Echidna concept of a discrete-stepping stick--slip mechanism with a small footprint, and requires no changes to the existing spine format.  Peak drive voltages have been reduced by more than an order of magnitude, to just \SI{9}{\volt} (see Table~\ref{tab:motor_summaries}).  A new control architecture has been designed and tested, made possible by these lower voltage demands.  This provides every actuator on every motor with a customisable drive waveform, meaning that each spine can now achieve its full performance potential.  Spines can also move in any direction, simplifying collision avoidance strategies.

Closed-loop positioning has been demonstrated that outperforms the existing Echidna technology in terms of both accuracy and speed.  The achieved RMS accuracy currently beats all other published methods, at \SI{< 2.8}{\micro\metre} in only five moves.  This has been shown to hold for \num{> 250000} positioning cycles over months of non-stop operation, proving that the design is robust.

Echidna technology's lowest possible tilt-induced throughput loss has been halved by achieving a $\num{\sim 1.4}\times$ increase of the current spine length limit, to \SI{\sim 354}{\milli\metre}.  This means that: a) survey signal-to-noise ratios improve, increasing scientific value; and/or b) spectrograph collimator speeds can be more efficiently matched to the nominal f-number of the instrument, reducing costs.  Good accuracy is maintained (\SI{4.0}{\micro\metre} RMS in six moves) and a further length increase looks plausible.

An additional avenue of investigation concerned fully modularising the control electronics to allow a distributed control system with stand-alone spine assemblies.  This appears to be feasible and would greatly reduce the size of a complete positioner system, although increases on-telescope heat output.

\begin{table}[!h]
    \caption{General motor characteristics for coarse and fine positioning modes with the existing Echidna and new designs. Note that \SI{5}{\micro\metre} step sizes can be achieved with the current design, but control limitations increase this for multiple spines for the reasons covered in this paper.  Values are approximate.}
    \centering
    \small
    \begin{tabular}{l l l}
        \hlinep
                                    & Existing motor            & New motor                     \\
        \hlinep
        Coarse mode step size       & \SI{40}{\micro\metre}     & \SI{40}{\micro\metre}         \\
        Coarse mode move speed      & \SI{4}{\milli\metre\per\second}
                                                & \SI{4}{\milli\metre\per\second}               \\
        Coarse mode voltage (pk)    & \SI{150}{\volt}           & \SI{9}{\volt}                 \\
        Fine mode step size         & \SI{10}{\micro\metre}     & \SI{5}{\micro\metre}          \\
        Fine mode voltage (pk)      & \SI{75}{\volt}            & \SI{3}{\volt}                 \\
        Maximum spine length        & \SI{250}{\milli\metre}    & \SI{\ge 354}{\milli\metre}    \\
        Technology readiness level (TRL)   & 9                  & 4                             \\
        \hlinep
    \end{tabular}
    \label{tab:motor_summaries}
\end{table}

Future development aims include: i) further prototyping to increase confidence in consistent performance; ii) testing under varying atmospheric conditions; iii) full design and prototyping of the stand-alone spine assembly concept; iv) further miniaturisation of the motor to achieve smaller pitches (e.g.\ \SI{< 7}{\milli\metre}, achieved previously with the existing technology \cite{2012SPIE.8446E..4WS}); and v) modelling of actuator hysteresis to improve open-loop performance for small moves when tracking objects.

This evolution of Echidna technology offers a quantum leap in terms of tilting spine fibre positioning performance and capabilities, without reinventing the wheel.

\nocite{2016thesis-gilbert-inpress}
\nocite{python}
\nocite{4160251}
\nocite{5725236}
\nocite{scipy}
\nocite{Hunter:2007}

\acknowledgments     

This work was funded by a Science and Technology Facilities Council (STFC) studentship.  Further assistance has been provided by the University of Oxford, the Australian Astronomical Observatory (AAO) and the Institution of Engineering and Technology (IET).  Attendance at SPIE Astronomical Telescopes + Instrumentation 2016 has been partially funded by an SPIE student travel grant.

\clearpage

\renewcommand{\refname}{REFERENCES}
\bibliography{refs}
\bibliographystyle{IEEEtran}

\end{document}